\documentclass[fleqn,twoside]{article}
\usepackage{amsmath,amssymb,espcrc2,color,cite}
\usepackage[usenames,dvipsnames]{xcolor}
\usepackage[percent]{overpic}

\usepackage{graphicx,soul}
\usepackage[figuresright]{rotating}
\usepackage{ulem}

\newcommand{\url}[1]{{\tt #1}}
\newcommand{\lsim}
{\;\raisebox{-.3em}{$\stackrel{\displaystyle <}{\sim}$}\;}

\newcommand{\ssi}{\ensuremath{\sigma^{\rm SI}_p}}

\newcommand{\ssdp}{\ensuremath{\sigma^{\rm SD}_p}}
\newcommand{\ssdn}{\ensuremath{\sigma^{\rm SD}_n}}
\newcommand{\Och}{\ensuremath{\Omega_\chi h^2}}

\newcommand{\tev}{\ensuremath{\,\, \mathrm{TeV}}}
\newcommand{\gev}{\ensuremath{\,\, \mathrm{GeV}}}

\def\refeq#1{\mbox{Eq.~(\ref{#1})}}

\def\reffi#1{\mbox{Fig.~\ref{#1}}}

\def\citere#1{\mbox{Ref.~\cite{#1}}}

\newcommand{\MC}{{\tt MasterCode}}
\newcommand{\MG}{{\tt MG5$\underline{~}$aMC@(N)LO}}
\newcommand{\MN}{{\tt MultiNest}}
\newcommand{\MO}{{\tt micrOMEGAs}}
\newcommand{\be}{\begin{equation}}
\newcommand{\ee}{\end{equation}}
\newcommand{\bea}{\begin{eqnarray}}
\newcommand{\eea}{\end{eqnarray}}

\newcommand{\gdm}{\ensuremath{g_{\rm DM}}}
\newcommand{\gsm}{\ensuremath{g_{\rm SM}}}

\def\approxprop{%
  \def\p{%
    \setbox0=\vbox{\hbox{$\propto$}}%
    \ht0=0.6ex \box0 }%
  \def\s{%
    \vbox{\hbox{$\sim$}}%
  }%
  \mathrel{\raisebox{0.7ex}{%
      \mbox{$\underset{\s}{\p}$}%
    }}%
}

\definecolor{orange}{rgb}{1,0.5,0}

\definecolor{Gray}{named}{Gray}

\graphicspath{{plots/}}

\title{\vspace{-4cm}
\bf \LARGE Global\,Analysis\;of\;Dark\,Matter\,Simplified\,Models\,with Leptophobic Spin-One Mediators using MasterCode \\ \vspace{0.5em}}

\author{
{\bf E.~Bagnaschi}\address[PSI]{Paul Scherrer Institut, CH--5232 Villigen, Switzerland},\address[DESY]{DESY, Notkestrasse 85, D-22607 Hamburg, Germany},
{\bf J.C.~Costa}\address[Imperial]
   {High\,Energy\,Physics\,Group,\,Blackett\,Laboratory,\,Imperial\,College,\,Prince\,Consort\,Road,\,London\,SW7\,2AZ,\,UK},
{\bf K.~Sakurai}\address[Warsaw]
{Institute of Theoretical Physics, Faculty of Physics, University of Warsaw, ul.~Pasteura 5, PL--02--093 Warsaw, Poland},
{\bf M.~Borsato}\address[HD]{Physikalisches Institut, Ruprecht-Karls-Universit\"at, D--69120 Heidelberg, Germany},
{\bf O.~Buchmueller}\addressmark[Imperial],
{\bf A.~De~Roeck}\address[CERNEP]
   {Experimental Physics Department, CERN, CH--1211 Geneva 23, Switzerland; \\  Antwerp University, B--2610 Wilrijk, Belgium},
{\bf M.J.~Dolan}\address[SLAC]
{ARC Centre of Excellence for Particle Physics at the Terascale, School of Physics, University of Melbourne, 3010, Australia},
{\bf J.R.~Ellis}\address[KCL]{Theoretical Particle Physics
  and Cosmology Group, Department of Physics, King's College London, London~WC2R~2LS, UK; \\
    National Institute of Chemical Physics and Biophysics, R{\" a}vala 10, 10143 Tallinn, Estonia; \\
Theoretical Physics Department, CERN, CH--1211 Geneva 23, Switzerland},
{\bf H.~Fl\"acher}\address[Bristol]
   {H.H.~Wills Physics Laboratory, University of Bristol, Tyndall Avenue, Bristol BS8 1TL, UK},
{\bf K.~Hahn}\address[NWU]
{Department of Physics \& Astronomy, Northwestern University, Evanston, Illinois 60208-3112, USA},
{\bf S.~Heinemeyer}\address[Madrid]
{Instituto de F\'{\i}sica Te{\'o}rica UAM-CSIC, C/ Nicolas Cabrera 13-15, E--28049 Madrid, Spain; \\
   Campus of International Excellence UAM+CSIC, Cantoblanco, E--28049 Madrid, Spain;\\
   Instituto de F\'{\i}sica de Cantabria (CSIC-UC), Avda. de Los Castros s/n,
    E--39005 Santander, Spain},
{\bf M.~Lucio}\address[USdC]{Instituto Galego de F{\' i}sica de Altas Enerx{\' i}as, Universidade de Santiago de Compostela, Spain},
{\bf D.~Mart\'inez~Santos}\addressmark[USdC],
{\bf K.A.~Olive}\address[Minnesota]
{William I.\ Fine Theoretical Physics Institute, School of Physics and
 Astronomy, University of Minnesota, Minneapolis, Minnesota 55455, USA},
{\bf S.~Trifa}\addressmark[Bristol],
{\bf G.~Weiglein}\addressmark[DESY]}

\begin{document}
\begin{abstract}
\vspace{0.25cm}

We report the results of a global analysis of dark matter simplified
models (DMSMs) with leptophobic mediator particles of spin one,
considering the cases of both vector and axial-vector interactions with
dark matter (DM) particles and quarks. We require the DMSMs
to provide all the cosmological DM density indicated by Planck and other
observations, and we impose the upper limits on spin-independent and
-dependent scattering from direct DM search experiments. We also impose
{all relevant LHC constraints from}
searches for monojet events and measurements of
the dijet mass spectrum. We model the likelihood functions for all the
constraints and combine them within the \MC\ framework, and
probe the full DMSM parameter spaces by scanning over the mediator and DM masses and couplings, {not fixing any of the model parameters.}
We find, in general, two allowed regions of the parameter spaces: one in
which the mediator couplings to Standard Model (SM) and DM particles may
be comparable to those in the SM and the cosmological DM density is
reached via resonant annihilation, and one in which the mediator
couplings to quarks are $\lesssim 10^{-3}$ and DM annihilation is
non-resonant. {We find that the DM and mediator masses may well lie within the ranges accessible to LHC experiments.}
We {also} present predictions for spin-independent and
-dependent DM scattering, and present specific results for ranges of the DM couplings
that may be favoured in ultraviolet completions of the DMSMs.
\vspace{0.25cm}
\begin{center}
{\tt KCL-PH-TH/2019-10, CERN-TH-2019-007, DESY-19-071, PSI-PR-19-06, IFT-UAM/CSIC-18-120\\
{FTPI-MINN-19/05, UMN-TH-3814/19}}
\end{center}
\end{abstract}
\thispagestyle{empty}
\newpage


\maketitle


\newpage

\section{Introduction}
\label{sec:intro}

The nature of Dark Matter (DM) is one of the most pressing issues in contemporary physics.
For over 80 years, astrophysical and cosmological observations have indicated its existence
indirectly~\cite{Zwicky,Rubin}. They have also suggested that DM is mainly cold, i.e., it has been non-relativistic since at least
the epoch of cosmological recombination~\cite{CDM}. Beyond this, we know very little, and candidates for DM
range in mass from black holes heavier than the Sun, to various species of (meta)stable neutral
particles, to very light particles with large occupation numbers best
described by classical fields{, see~\citere{review} for a review}.
If the DM is composed of weakly-interacting massive particles (WIMPs) that were in thermal equilibrium with Standard Model (SM) particles in the early
Universe, freeze-out calculations suggest that the WIMP is likely to weigh ${\cal O}$(TeV), in which case
it could be produced at accelerators, notably the LHC. Thus, the search for WIMP DM particles has been
one of the principal research objectives of the LHC experiments,
particularly ATLAS and CMS~{\cite{Abdallah:2015ter,Abercrombie:2015wmb,Boveia:2016mrp,Albert:2017onk,Abe:2018bpo}}.

Several approaches have been taken to DM searches at the LHC, in both predictions for possible
signals and interpretations of the {search limits}. Initially, many analyses were
based on specific models that predict WIMPs capable of
providing the cold DM, such as supersymmetry (SUSY)~\cite{EHNOS} and some other models with new physics at the TeV scale~\cite{LKP}.
These searches were typically either for the production of heavier new particles
followed by cascade decays to the dark matter particle, or direct production of dark matter particles in association
with a single SM particle used for tagging purposes (the 'mono-X' signature).
{DM models giving rise to long-lived signatures have also been studied recently~\cite{LLDM}.}

We have used the \MC\ framework~\cite{mcweb} {{previously} to study several SUSY scenarios
with various experimental signatures~\cite{MC}, and other groups have made
parallel analyses~\cite{others}.
{In particular, in \citere{MC} we provided detailed predictions for
  the preferred parameter spaces in various concrete SUSY models.}
However, there are many other scenarios for possible physics beyond the SM,
{and it is desirable to fashion search strategies that facilitate the interpretation of results
with differing characteristics, minimizing theoretical biases.}

Recently, growing attention is being paid to more general
model frameworks and analyses motivated by more generic experimental signatures. One of the
first such approaches was to construct an effective field theory (EFT) for the interactions between
DM and SM particles~\cite{Goodman:2010yf,Goodman:2010ku,Zheng:2010js,Fox:2011pm}, with a focus on the mono-X signatures.
A shortcoming of this EFT approach is that, in order to obtain the
appropriate cosmological DM density, the DM/SM interactions are likely
to be mediated by particles in the TeV mass range, for which the approximation of a contact
interaction may be inadequate. In particular, the LHC might also be capable of producing the mediator
particle directly, providing an additional signature beyond the EFT framework~\cite{Busoni:2013lha,Buchmueller:2013dya}.
For this reason, interest has been growing in Dark Matter Simplified Models (DMSMs)~\cite{Buchmueller:2013dya,Buchmueller:2014yoa,Papucci:2014iwa,Harris:2014hga,Buckley:2014fba,Kahlhoefer:2015bea,1403.4837,Bell:2015sza,Fairbairn:2016iuf},
which postulate effective Lagrangians that include explicitly the mediator particle and its interactions with
both DM and SM particles. Recommendations for the formulation of DMSMs for use by the LHC
experiments have been made by the LHC Dark Matter Working Group (DMWG)~\cite{Abdallah:2015ter,Abercrombie:2015wmb,Boveia:2016mrp,Albert:2017onk,Abe:2018bpo}.

In this paper we report the results of global likelihood analyses within the \MC\ framework of {some of the} DMSMs
suggested by the DMWG and analyzed by ATLAS and CMS.

We take a bottom-up approach, postulating a mediator particle in the LHC
mass range, without requiring an explicit UV completion, {commenting later how} the
models investigated here could arise in UV-complete scenarios.
We consider only $s$-channel mediators of spin one.
Avoiding the strong LHC constraints on mediators interacting with charged leptons~\cite{Aaboud:2017buh,Sirunyan:2018exx}, we
assume that their interactions are leptophobic.  {We assume that the DM particle is a neutral Dirac fermion $\chi$. 
The Lagrangian for the spin-one mediator $Y$ interactions with the DM particle $\chi$ is
\be
{\cal L}^1_\chi \; = \; \bar{\chi} \gamma_\mu \left( g_\chi^V + g_\chi^A \gamma_5 \right) \chi Y^\mu \, ,
\ee
and that for its interactions with quarks is
\be
{\cal L}^1_q \; = \; \sum_{i} \bar{q_i} \gamma_\mu \left( g_{q_i}^V + g_{q_i}^A \gamma_5 \right) q_i Y^\mu \, .
\ee
For simplicity, following the DMWG~\cite{Albert:2017onk} we assume that the interactions between the $Y$-boson and quarks are generation-independent,
and we also assume that they are the same for charge-2/3 and -1/3 quarks ($g_q^{V,A} := g_{q_i}^{V,A} \;\forall\; i$)~\footnote{We do not discuss here the cancellation of gauge anomalies, which
is possible with a suitable non-minimal `dark sector', whose experimental signatures are more model-dependent
and  not discussed in detail here~\cite{Ismail:2016tod,Ellis:2017tkh,Ellis:2018xal}.}}.
Being leptophobic,
the $Y$ boson has no interactions with leptons.
We consider two scenarios for the $Y$ couplings,
one in which they are purely vectorial:
\bea
g_\chi^V \; \equiv \; \gdm \; \ne \; 0, & & g_\chi^A \; = \; 0 \, , \nonumber \\
g_q^V \; \equiv \; \gsm \; \ne \; 0, & & g_q^A \; = \; 0 \, ,
\label{YV}
\eea
and another in which they are purely axial:
\bea
g_\chi^A \; \equiv \; \gdm \; \ne \; 0, & & g_\chi^V \; = \; 0 \, , \nonumber \\
g_q^A \; \equiv \; \gsm \; \ne \; 0, & & g_q^V \; = \; 0  \, .
\label{YA}
\eea
Note that the axial vector model could in principle violate unitarity at high energies. However, the couplings required
for this to occur are larger than the range we consider in our scans, and would even call the particle interpretation of the
mediator into question~\cite{Kahlhoefer:2015bea,Englert:2016joy}.

{UV completions of the DMSMs we study could} be achieved in a number of different ways.
{These could favour specific ranges of \gsm, \gdm\ and their ratio, as we explore below.}
Since our spin-1 models involve massive vector bosons, they will likely involve a dark Higgs responsible for giving them masses,
or else a St\"uckelberg mechanism~\cite{Bell:2016uhg}.
However, our intention
here is not to fit the parameters of a complete model, but rather to study correlations between the model parameters defined above,
which are the minimal sets leading to an interplay between LHC and DM
phenomenology that is of interest.

Each of these DMSMs has four free parameters, the masses of the DM
particle and the mediator, $m_\chi$ and $m_{Y}$,
and the mediator couplings to the DM and SM particles, $\gdm$ and
$\gsm$. Sampling these 4-dim} parameter spaces is
computationally tractable. However, experimental results are often interpreted
for fixed values of the couplings, and thus {previous investigations
have generally fixed} two out of the four parameters. A more general
approach is desirable for combining the direct DM constraints with those from the LHC, particularly because obtaining the
preferred cosmological value of \Och\ requires values of $\gdm$ and $\gsm$ that depend on $m_\chi$ and {$m_{Y}$}.
{Consequently, without fixing any parameter of the model,} we implement the predictions of the models using {\tt \MO}~\cite{Micromegas} and
{\tt Madgraph5\_aMC@NLO}~\cite{MG} using {\tt DMSIMP} model files~\cite{DMSIMP} to
build a likelihood function for each of the astrophysical and
accelerator constraints, and combine them using the \MC\ framework. We use \MN {\cite{multinest}} to identify the regions where the global likelihood is minimized.
{We do not include in our analysis indirect searches for astrophysical DM particles via their annihilations
into SM particles, which were found in~\cite{Ellis:2018xal} to be unimportant in simple DM models {when} $m_\chi \gtrsim 50 \gev$.}

The layout of this paper is as follows. In Section~\ref{sec:constraints} we set out the constraints we use,
and in Section~\ref{sec:framework} we describe our analysis framework. Our results are described in
Section~\ref{sec:results}, Possible UV completions are discussed in Section~\ref{sec:gdmeqgsm}, and our conclusions and some perspectives are set out in Section~\ref{sec:conx}.


\section{Astrophysical and LHC Constraints}
\label{sec:constraints}

\subsection{Dark Matter Constraints}

{\subsubsection{Dark Matter density}}

The mean density of cold DM in the Universe is tightly constrained by Planck measurements of the cosmic
microwave background and other observations~\cite{Planck}:
\be
\Omega_{\rm CDM} h^2 \; = \; 0.120 \pm 0.001 \, .
\label{OmegaCDM}
\ee
We assume here that the dominant source of this cold DM is the WIMP $\chi$, so that
$\Och \simeq \Omega_{\rm CDM} h^2$, with interactions that are described
in the Lagrangians (\ref{YV}) and (\ref{YA}) in the DMSMs we study. 

{The relic abundance is approximately proportional to the inverse of the annihilation cross-section, 
$\Omega_\chi h^2 \propto 1/\langle \sigma v \rangle$.
If the annihilation is dominated by the $s$-channel process $\chi \chi \to Y^{(*)} \to$ Standard Model particles, 
the leading expression for the cross-section takes the form \cite{Arcadi:2014lta,imy}
\begin{equation}
\langle \sigma v \rangle_s \simeq 
\langle
\frac{c_Y}{32 \pi} \frac{\gsm^2 \gdm^2 m^2_\chi}{(m_\chi^2 v^2 + 4 m_\chi^2 - m_Y^2)^2 + m_Y^2 \Gamma_Y^2} \rangle\,,
\label{eq:s-xs}
\end{equation}
where $c_Y$ is an ${\cal O}(1)$ constant for the vector mediator, whereas for the 
axial mediator it is suppressed by the quark mass and the relative velocity, $v$, of the 
annihilating DM particles: $c_Y \sim a \frac{m_q^2}{m_\chi^2} + b v^2$.
The total width of $Y$ is given by $\Gamma_Y = \Gamma_Y^{\chi \chi} + N_c \sum_q \Gamma_Y^{qq}$, 
where $N_c = 3$ is a colour factor and
\begin{equation}
\Gamma_Y^{XX} \; = \; \frac{g_X^2 m_Y}{12 \pi} \cdot f \left( \frac{m_X^2}{m_Y^2} \right) \, ,
\end{equation}
where $X = \chi \, ({\rm DM}), q \, ({\rm SM})$, and
\begin{eqnarray}
f(x) 
\; = \; \begin{cases}
    (1 + 2 x) (1 - 4 x)^{1/2}  & ({\rm vector}) \\
    (1 - 4 x)^{3/2}  & ({\rm axial\,vector}) 
  \end{cases}\, ,
\end{eqnarray}
{where $x \equiv (m_X/m_Y)^2$.}
The width of $Y$ becomes particularly important in the resonant region, where $m_Y \simeq 2 m_\chi$ 
and $\Gamma_Y^{\chi \chi} \ll \Gamma_Y$} {because of phase-space suppression}.

{In the axial-vector case, $\langle \sigma v \rangle_s$
is approximately proportional to $v^2$ through $c_Y$
and, if $\gsm \ll 1$, the velocity term dominates the denominator of
Eq.~\eqref{eq:s-xs}
at the resonance. This leads to the approximate scaling
\begin{equation}
\Omega_\chi h^2 \approxprop \frac{1}{\langle \sigma v \rangle_s} 
\approxprop \frac{m_Y^2}{\gsm^2 \gdm^2} \, ,
\label{Omegagdm_axial}
\end{equation}
{where $\approxprop$ denotes approximate proportionality}. On the other hand, in the vector case 
the annihilation is $s$-wave and
}
the narrow-width approximation is applicable for $\gsm \ll 1$, and 
one finds that in the limit $\gsm \to 0$
\begin{align}
&  1/[(m_Y^2 - 4 m_\chi^2)^2 + m_Y^2 \Gamma_Y^2] \nonumber \\
& \qquad \qquad \qquad \to \delta(m_Y^2 - 4 m_\chi^2)/(m_Y \Gamma_Y) \, ,
\end{align}
so the scaling of the relic density is expected to be
\begin{equation}
\Omega_\chi h^2 \approxprop \frac{1}{\langle \sigma v \rangle_s} 
\approxprop \frac{m_Y^2}{\gdm^2}
\label{Omegagdm}
\end{equation}
near the $Y$ peak. However, we recall that $\chi \chi$ annihilation takes place in a thermal bath, with collisions taking place at a range of centre-of-mass energies $\ge 2 m_\chi$. For this reason, annihilations at an effective
centre-of-mass energy within $\mathcal{O}(\Gamma_Y)$ of $m_Y$
do not dominate when $\Gamma_Y$ is very small. We find numerically that (\ref{Omegagdm}) captures the \gdm\ dependence of $\Omega_\chi h^2$, and that it depends only weakly on \gsm\ over an intermediate range of \gsm,
but that $\Omega_\chi h^2$ increases at both large and small \gsm\ {:
\begin{equation}
\Omega_\chi h^2 \; \approxprop \; \frac{m_Y^2 \gsm^2}{\gdm^2} 
\label{gsm>gdm}
\end{equation}
for $\gsm \gg \gdm$ and
\begin{equation}
\Omega_\chi h^2 \; \approxprop \; \frac{m_Y^2}{\gsm^2}
\label{gsm<gdm}
\end{equation}
for $\gsm \ll \gdm$. 
The dependence on $\gdm$ in this case depends on the relative sizes of the width and mass difference between $m_Y$ and $2 m_\chi$ \cite{imy}. If $\Gamma_Y/m_Y > (1 - 4m_\chi^2/m_Y^2)$ and  $m_Y > 2 m_\chi$ then $\Omega_\chi h^2$ is
proportional to $\gdm^2$. However, if the width is very small (for example if $\gsm$ is very small and $m_Y < 2 m_\chi$), then the denominator in
(\ref{eq:s-xs}) is dominated by the mass difference and $\Omega_\chi h^2$ is
inversely proportional to $\gdm^2$, {as seen in Eq. (\ref{Omegagdm_axial}) for the axial-vector case}.

Another important process is the $t$-channel process $\chi \chi \to Y Y$ for $m_\chi > m_Y$.  
The leading contribution to the $t$-channel cross-section is given by \cite{Arcadi:2014lta}
\begin{equation}
    \langle \sigma v \rangle_t
    \simeq
    \frac{\gdm^4}{32 \pi} \frac{(m_\chi^2 - m_Y^2)^{3/2} }{m_\chi (2 m_\chi^2 - m_Y^2 )^2} \,,
    \label{eq:Oh2_t}
\end{equation}
and is kinematically forbidden when $m_\chi < m_Y$.

We use \MO~\cite{Micromegas} to evaluate \Och\ {numerically}
in each model parameter set, {including a ${\pm 10}$\% range around \refeq{OmegaCDM}}. {This enables us to obtain a substantial but unbiased sample
of model points.

{As a first validation step, we show in \reffi{fig:Och}}
a comparison between our \MO\
calculations of \Och\ (colours and solid blue line) with the $\Och = 0.12$ curve shown by the CMS Collaboration in~\cite{CMS-EXO-048}
(dashed line) for the case of a spin-one mediator with vector-like couplings $\gsm^V = 0.25, \gdm^V = 1$.
As seen in \reffi{fig:Och}, the agreement is excellent, providing a valuable cross-check on our calculations. We note that the
red shaded region at larger values of $m_Y$ is excluded because $\Och > 0.12$, and that the DM particle is
underdense in the blue region at large $m_\chi$.

\begin{figure}[]
\centering
\includegraphics[width=0.475\textwidth]{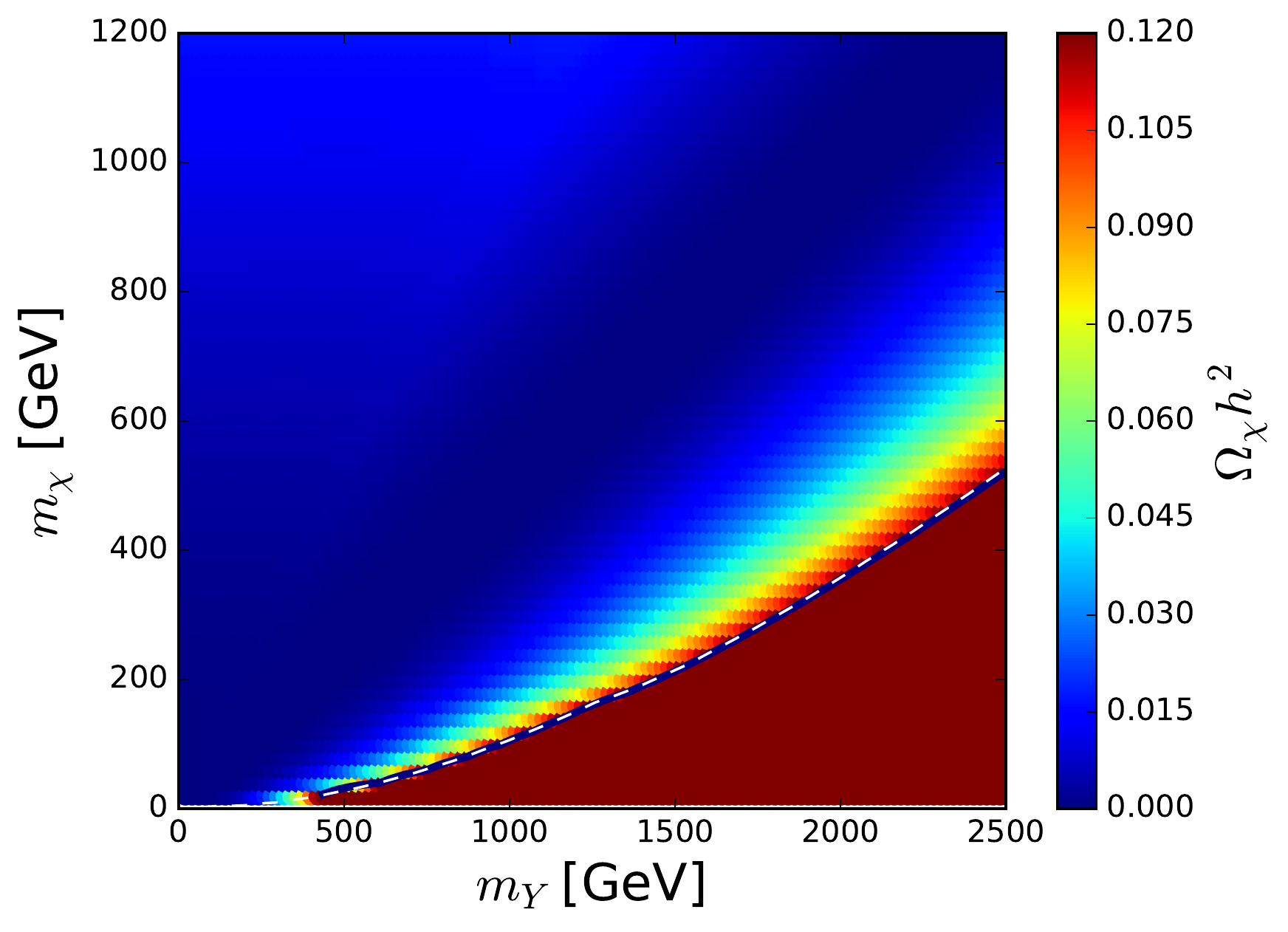}
\caption{\it Comparison of our calculation of \Och\ (colours and solid blue line) with the $\Och = 0.12$
curve shown by the CMS Collaboration in~\cite{CMS-EXO-048} (dashed white line) for the DMSM with a spin-one mediator having
vector-like couplings $\gsm^V = 0.25$, $\gdm^V = 1$ to quarks and to DM
particles $\chi$, respectively.}
\label{fig:Och}
\end{figure}

{\subsubsection{Spin-independent DM scattering}}

We evaluate the constraint on the spin-independent
DM scattering cross-section $\ssi$
from a combination of the LUX~\cite{LUX}, PandaX-II~\cite{PANDAX} and XENON1T~\cite{XENON} experiments using a joint likelihood function,
evaluating the theoretical prediction for $\ssi$ using \MO.
{This constraint is important for the vector mediator case, where the spin-independent cross-section has the following scaling \cite{Boveia:2016mrp} $\ssi \propto \gdm^2 \gsm^2/m_Y^4$, with the following approximate numerical value:
\begin{equation}
    \ssi \simeq  
    \left( \frac{\gdm}{0.1} \right)^2 
    \left( \frac{\gsm}{0.1} \right)^2
    \left( \frac{m_Y}{1\,\rm TeV} \right)^{-4} 
    10^{-43} [{\rm cm}^{-2}]
    \label{eq:ssi_expression}
\end{equation}
in the $s$-channel region where $m_Y \simeq 2 m_\chi$.
The $\ssi$ constraint provides important limits on the vector mediator mass $m_Y$ and the product $\gdm \gsm$.}

{We allow for an overall uncertainty of 10\% in the spin-independent cross section.}

{\subsubsection{Spin-dependent DM scattering}}

We also evaluate the constraint imposed by the upper limit on the spin-dependent
DM-proton scattering cross-section $\ssdp$ from the PICO-60 experiment~\cite{PICO},
and that on the spin-dependent DM-neutron scattering cross-section $\ssdn$ from the XENON1T experiment~\cite{XENONspindep}.
We evaluate theoretical predictions for $\ssdp$ and $\ssdn$ using \MO, {which uses estimates of the spin-dependent scattering
matrix elements consistent with}~\cite{ENODM}.

{We also allow for an overall uncertainty of 10\% in the spin-dependent cross section.}

\subsubsection{Halo model}

The majority of direct searches for DM scattering assume a standard halo model (SHM) in which the local density of cold DM is 0.3~GeV/cm$^3$,
{with negligible uncertainty. However, several recent analyses favour a larger local density, and here we follow~\cite{McCabe} in assuming a central value for
the local DM density of 0.55~GeV/cm$^3$. {For this reason, the direct DM scattering limits, estimated future sensitivity curves and neutrino `floors' we use are rescaled relative to the published curves.}}

{It has been suggested in a recent analysis
of Gaia data that there may be a local debris flow that modifies the local velocity distribution from that given by the SHM~\cite{Lisanti}. However, this
modification of the the local velocity distribution was shown to
have only a small effect on the interpretation of the XENON1T experiment for $m_\chi$ in the range of interest in this paper,
{and a similar conclusion was reached in~\cite{McCabe}.}
For another estimate of the local density of DM based on Gaia data, see~\cite{newGaia}.}

{When implementing the experimental constraints on $\ssi$ and $\ssdp, \ssdn$, we assume an overall uncertainty of 30\%~\cite{LocalDMdensity,McCabe},
in the local density.}


\subsection{LHC Experimental Constraints}

\subsubsection{Monojet Constraints}

{
We use the constraints provided by the CMS Collaboration in~\cite{CMS-EXO-048}
using 35.9/fb of data from collisions at 13~TeV, {based on an analysis
employing signal regions targeting monojet final states.
In particular, we use the 95\,\% CL upper bounds $R_i^{\rm UL}({\bf m})$,
where $i$ labels the vector and axial-vector
mediators,
imposed on $\sigma({\bf m}) / \sigma^{\rm fix}({\bf m})$ evaluated at each point in the ${\bf m} = (m_\chi, m_Y)$ plane,
where $\sigma^{\rm fix}$ is the cross-section evaluated with $(\gsm, \gdm) = (0.25, 1)$.

We model the likelihood function from this
search using the procedure outlined in {\tt Fastlim}~\cite{Fastlim}:
\be
\Delta \chi^2 \; = \; 5.99 \times \left(
\frac{1}{ R_i^{\rm UL}({\bf m}) } \frac{\sigma_{\rm MG}({\bf m})}{\sigma^{\rm fix}_{\rm MG}({\bf m})} \right)^2 \, ,
\label{monojetlimit}
\ee
where $\sigma_{\rm MG}$ is the monojet cross-section calculated using \MG~\cite{MG}.

As a second step in our validation, we {display in \reffi{fig:monojet}
our implementations (colours and solid red lines) of the CMS monojet constraints on
the spin-1 mediator with vector-like interactions (left panel) and axial
couplings (right panel). We find excellent agreement with the CMS limits~\cite{CMS-EXO-048},
{which are indicated by dashed red lines}.
}
\begin{figure*}[]
\centering
\includegraphics[width=0.475\textwidth]{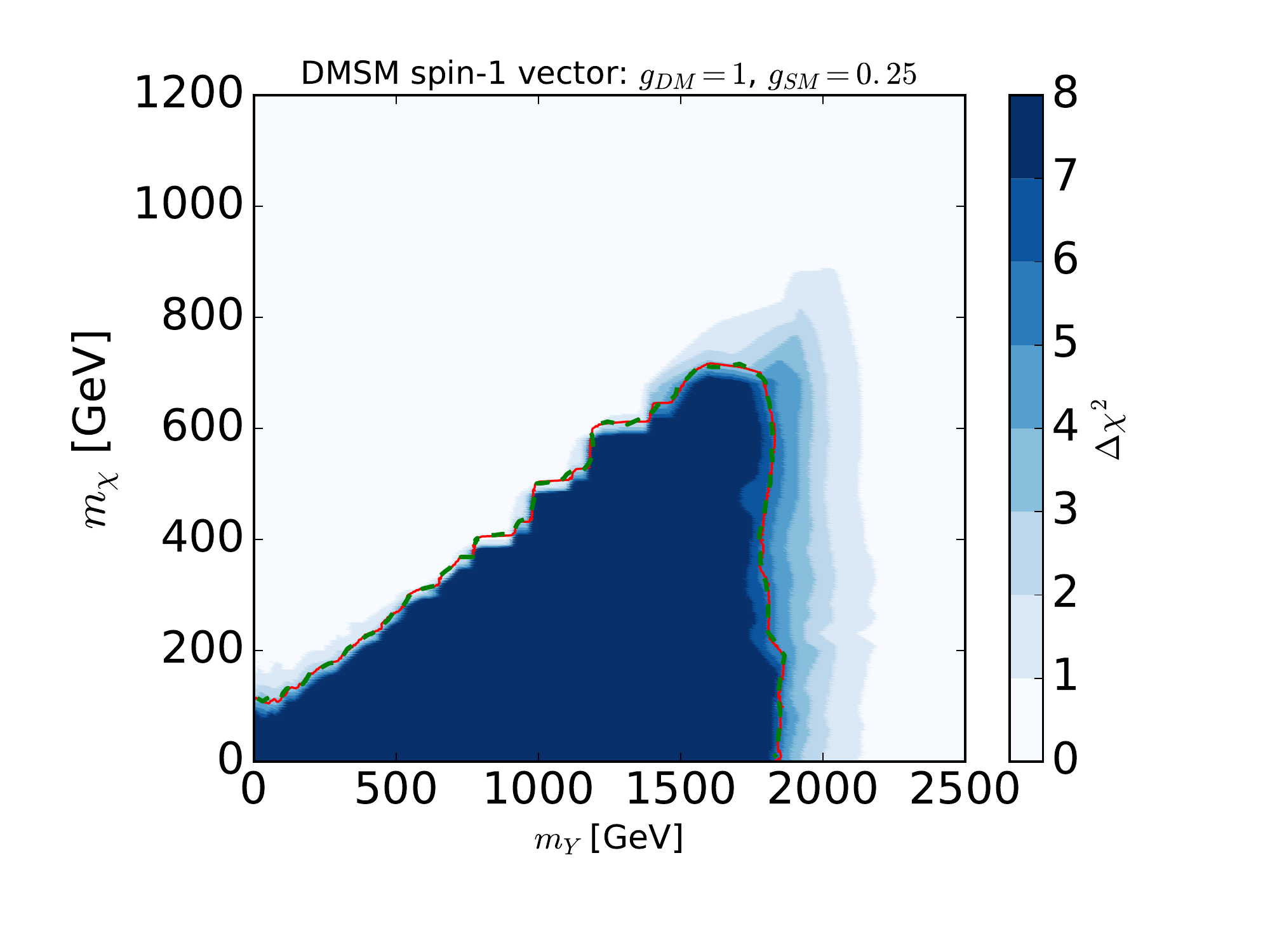}
\includegraphics[width=0.475\textwidth]{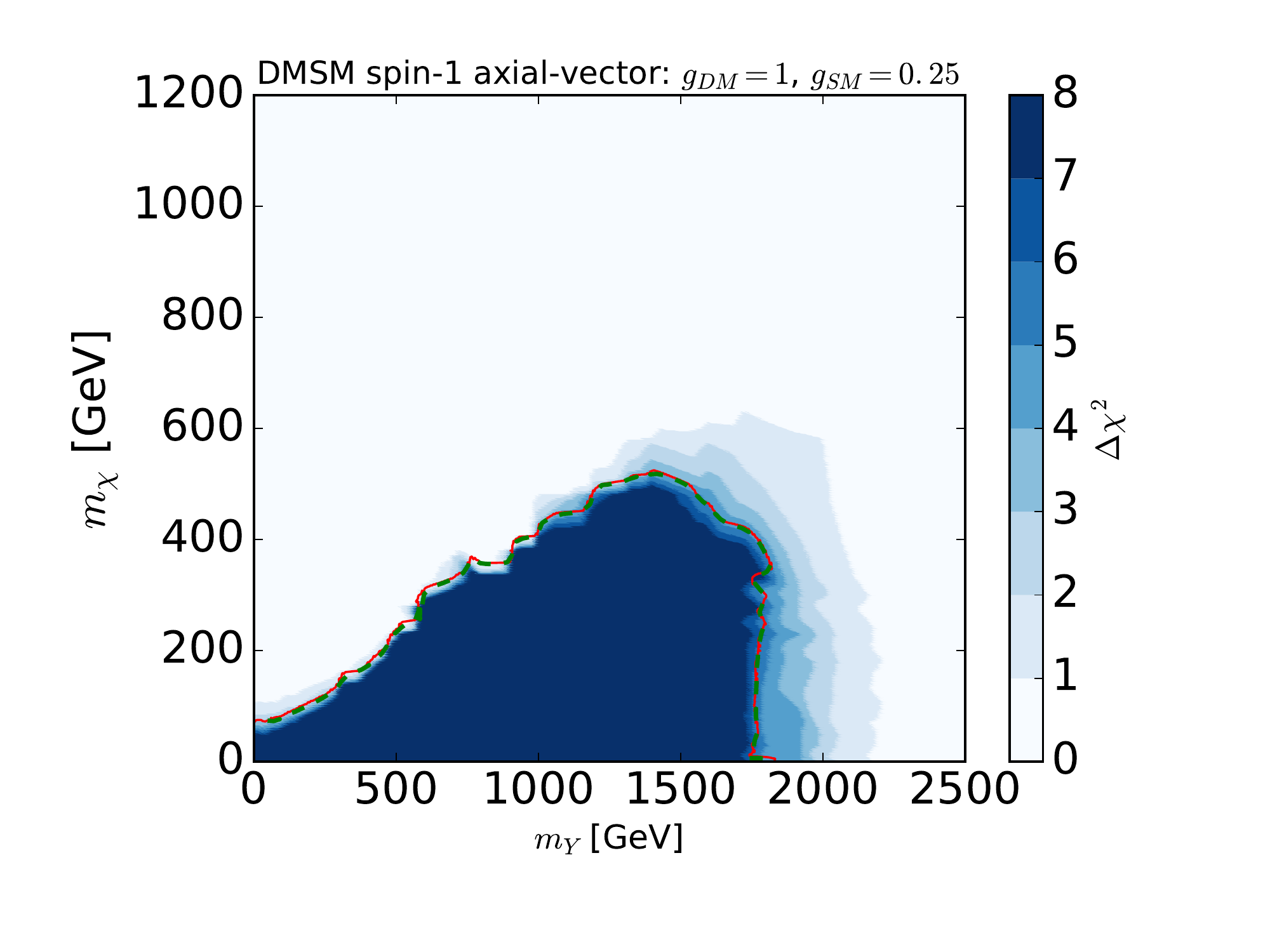}
\caption{\it The likelihoods for our implementations (\protect\ref{monojetlimit})
(colours and solid red lines) of the CMS monojet constraints~\protect\cite{CMS-EXO-048}
{(dashed green lines)} for a spin-one mediator having (left panel)
vector-like couplings and (right panels) axial couplings to quarks and
DM particles.}
\label{fig:monojet}
\end{figure*}


\subsubsection{Dijet Constraints}

The ATLAS and CMS Collaborations have both published constraints {from dijet invariant-mass distributions}
on $Z'$ resonances~\cite{CMS-PAS-EXO-16-056,ATLAS-CONF-2016-070,ATLAS-TLA-dijet,CMS-JHEP-01-2018-097,ATLAS-PRD96-052004,1802.06149}
{with up to 139/fb of data at 13~TeV~\cite{ATLAS:2019bov}.}
We use {a combination of} the limits presented in the plane of the $Z'$ mass and its ${\bar q} q$ coupling, $g_q$.
In recasting this constraint for the $pp \to Y \to qq$ process in our DM models,
we demand
\be
\sigma_{Y \to qq} \; \le \; \sigma(g_q^*)
\label{dijetcondi}
\ee
at $m_Y = m_{Z'}$, where $g_q^*$ is the upper limit placed
{on the coupling of the $Z'$ to quarks}
at a given $m_{Z'}$.
We note that $\sigma_{Y \to qq}$ and $\sigma({g_q^*})$
can be written as {(using $\Gamma_x \equiv \Gamma(Y \to xx)$)
\be
\sigma_{Y \to qq} = c \, \frac{\gsm^4}{\Gamma_{q} + \Gamma_{\chi}},
\quad
\sigma(g_q^*) = c  \, \frac{(g_q^*)^4}{\Gamma^*_{q}}
\ee
} with the same constant $c$ at $m_Y = m_{Z'}$,
and
$\Gamma_q = \big( \frac{\gsm}{g_q^*} \big)^2 \Gamma_q^*$.
Using these relations, the equality in Eq.~\eqref{dijetcondi} occurs when {\gsm\ takes the value}
\be
\gsm^* \; \equiv \; \sqrt{\frac{(g^*_q)^2}{2} \left( 1 + \sqrt{1 + \frac{4 \Gamma_{\chi}}{\Gamma^*_q}} \right) } \, ,
\ee
which serves as the upper bound on
$\gsm$ for given $m_Y$, $\Gamma_q$ and $\Gamma_{\chi}$.
We model the corresponding likelihood function using
\be
\hspace{-0.1cm}
\Delta \chi^2  = 4 \times
\left(
\left[ \frac{\gsm^4}{\Gamma_{q} + \Gamma_{\chi}} \right]
/
\left[ \frac{(\gsm^*)^4}{\Gamma_{q}(\gsm^*) + \Gamma_{\chi}} \right]
\right)^2
\ee
where $\Gamma_{q}$ and $\Gamma_{\chi}$ are obtained with \MG.
}

{As a further step in our validation, we}
have cross-checked our recasting with the
CMS and ATLAS limits~\cite{CMS-PAS-EXO-16-056,ATLAS-CONF-2016-070,ATLAS-TLA-dijet,CMS-JHEP-01-2018-097,ATLAS-PRD96-052004,1802.06149,ATLAS:2019bov},
as shown in Fig.~\ref{fig:lyingTrump}~\footnote{{The discontinuity in the limit at $m_Y = 2 \tev$ is due to the improved
sensitivity of the recent ATLAS limit using 139/fb of data at 13~TeV~\cite{ATLAS:2019bov}.}}.
{The coloured lines correspond to the various experimental analyses,
which can directly compared to the \MC\ $\chi^2$ evaluation indicated by
{colours in the} plane.}
Note that we use the most sensitive upper limit for each mass value, and do not attempt
to combine different experiments. Since the width of the {band} in Fig.~\ref{fig:lyingTrump} where the
{variation in the likelihood function  is significant} {(as indicated by the colour transition)}
is typically smaller than the differences between the various experimental limits,
this should be a good approximation. The excellent agreement between our 95\% CL upper limit on $\gsm$
and that quoted by the experiments is a valuable cross-check of our implementation of the dijet invariant-mass constraints.

\begin{figure}[]
\centering
\includegraphics[width=0.5\textwidth]{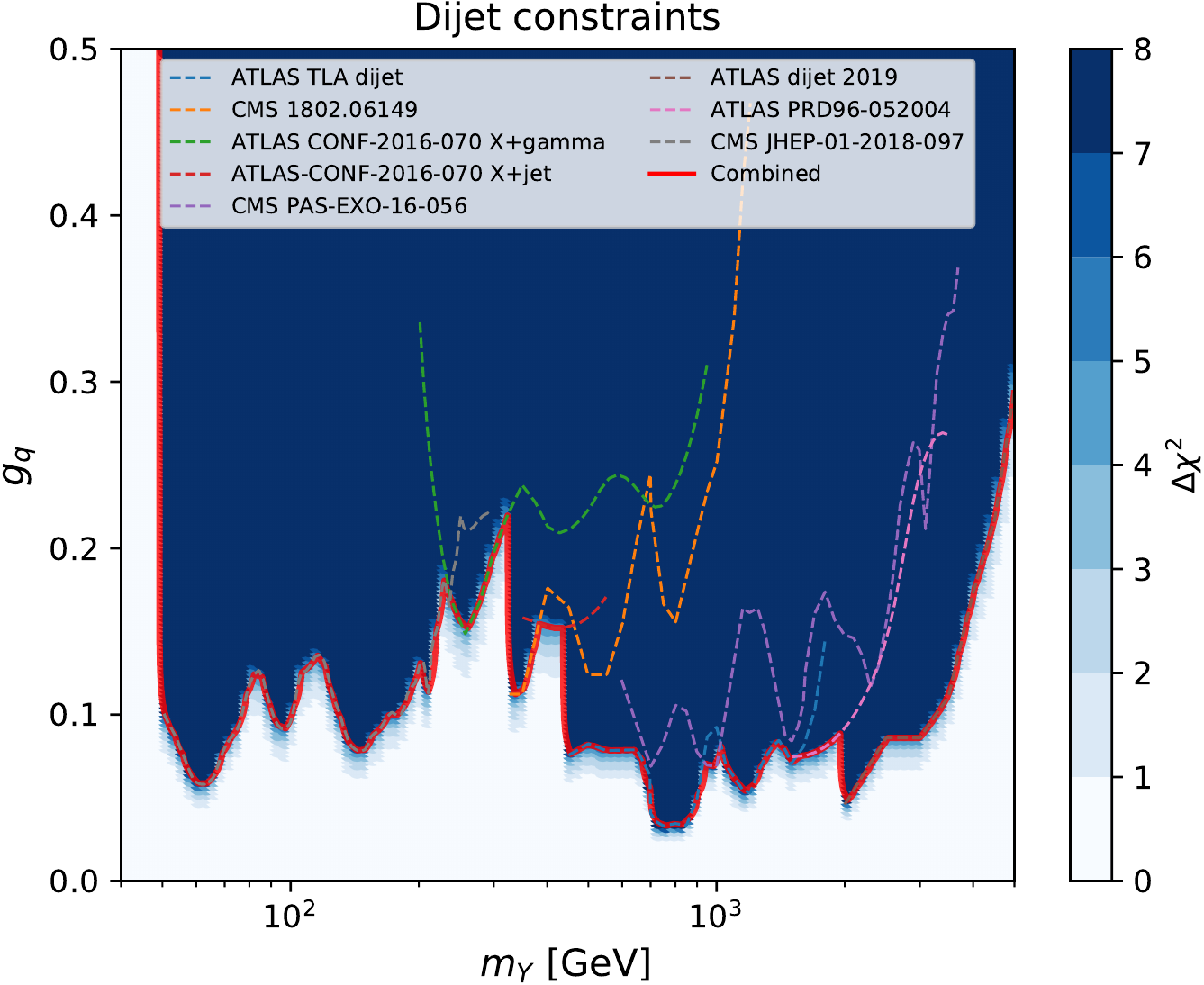}
\caption{\it Comparison of our implementation of the LHC dijet {invariant-mass} constraint, {indicated by the colour coding,} with a compilation
of the limits published by the ATLAS and CMS Collaborations~\protect\cite{CMS-PAS-EXO-16-056,ATLAS-CONF-2016-070,ATLAS-TLA-dijet,CMS-JHEP-01-2018-097,
ATLAS-PRD96-052004,1802.06149,ATLAS:2019bov}, for a $Z^{'}$ boson, in the plane of its mass and coupling to the quarks.}

\label{fig:lyingTrump}
\end{figure}

{In addition to these invariant-mass searches, the CMS Collaboration has also published a limit on $\gsm$
as a function of $m_{Z'}$ from an analysis of dijet angular distributions~\cite{CMSangle}. {However, this limit is weaker than the 
limits from the dijet invariant mass distribution for the range of masses we study.}
Accordingly, we do not include it in our analysis.}


\section{Analysis Framework}
\label{sec:framework}

Global-fitting frameworks such as \MC~\cite{MC} have been used to study the constraints on different possible extensions of the SM,
including SUSY models in particular, but can also be applied to other models with WIMP candidates for DM.
Since DMSMs are not complete models, we do not consider it meaningful to compute $p$-values.
Instead, here we use the \MC\ framework to correlate the impacts of constraints from different sectors,
explore correlations and find regions of the model parameters that are still allowed.

We recall that \MC\ is a frequentist framework for combining constraints, written in {\tt Python/Cython} and {\tt C++}.
As already mentioned, we use \MG~\cite{MG} to calculate mediator properties and the collider constraints on models
implemented using {\tt DMSIMP}~\cite{DMSIMP},
and we use \MO~\cite{Micromegas} for DM density and scattering calculations.
We use the \MN {~\cite{multinest}} algorithm for efficient sampling of the model parameter spaces. Since each of the DMSMs
(\ref{YV}) and (\ref{YA}) that we study has a parameter space of only 4 dimensions, and since
the constraint set is not large, sampling the model parameter spaces is not computationally onerous.
{In studies for our analysis we have made use of} {\tt udocker} software framework~\cite{udocker}
to automatize the deployment of \MC\ inside Linux containers.
This is a middleware suite developed in the context of the {\tt INDIGO} datacloud project~\cite{INDIGO} to run
{\tt docker} containers in userspace, without requiring root privileges for installation or for execution.

The ranges of DMSM parameters that we study are shown in Table~\ref{tab:ranges},
together with the numbers of segments we use {for our basic sampling of the parameter space}.
{The range of $m_Y$ was chosen to avoid the low-mass region where mixing with the $Z$
could be {subject to important constraints from precision electroweak data}~\footnote{{See~\cite{Ellis:2018xal} for
a treatment of these constraints, which we discuss in more detail below.}} and indirect searches for astrophysical DM annihilations should be considered~\footnote{{Searches for $\gamma$-rays from hadronic DM annihilations are generally insensitive to the 
cross-section required to obtain the correct cosmological DM density for $m_\chi \gtrsim 50 \gev$~\cite{Ellis:2018xal}, see also~\cite{FLAT} and references therein. Indirect constraints from searches for energetic solar neutrinos are not competitive with direct searches for spin-dependent DM scattering, as discussed below.}},
but} include all the masses for which LHC searches are sensitive. 
The couplings $\gsm$ and $\gdm$ were restricted to perturbative ranges {$< \sqrt{4 \pi}$}.

\begin{table}[htb!]
\begin{center}
\resizebox{0.475\textwidth}{!}{
\begin{tabular}{|c|c|c|}
\hline
Parameter   &  \; \, Range      & \# of Segments \\
\hline \hline
${m_Y}$     &  $(0.1, 5) \tev$  & 10 \\
$m_\chi$       &  $(0, 2.5) \tev$  & 8 \\
$\gsm$       &  $(0, \sqrt{4 \pi})$  & 2 \\
$\gdm$       &  $(0, \sqrt{4 \pi})$  & 2 \\
\hline
\multicolumn{2}{|c|}{Total \# of segments}& 320 \\
\hline
\end{tabular}}
\caption{\it The ranges of the DMSM parameters sampled, together with the numbers of
segments into which they were divided {during the sampling}.}
\label{tab:ranges}
\end{center}
\end{table}

{In addition to the generic sampling ranges shown in Table~\ref{tab:ranges}, we have made dedicated scans of
certain regions in order to clarify certain DMSM features. In particular, we gathered dedicated samples
of the regions where $2m_\chi/m_Y$ deviates from unity by $< 10^{-3}$, so as to sample adequately annihilations near the $Y$ peak when $\Gamma_Y \ll m_Y$. We used similar sampling procedure for both the vector and axial-vector DMSMs,
generating $\sim 100$~million parameter sets in each case.}

\section{Results}
\label{sec:results}

\subsection{Vector DM Couplings}
\label{sec:vector}

Fig.~\ref{fig:vectormassplane} displays the $(m_Y, m_\chi)$ plane in the vector-like model (\ref{YV})
after application of the constraints discussed above. The parameter regions
with $\Delta \chi^2 < 2.30$ (5.99), which are favoured at the 68\% (95\%)~CL
and regarded as proxies for 1- (2-)$\sigma$ regions, are delineated by red (blue)
contours, respectively. {In this and subsequent figures, 
we illustrate the dominant mechanism bringing the DM density into the allowed range at the point that minimizes $\chi^2$ in the displayed 2-dimensional projection of the four-dimensional parameter space using colour coding}: {\\
$\bullet$~Green: annihilation via $t$-channel
$\chi$ exchange into pairs of mediator particles $Y$ that subsequently decay into SM particles, {in the region where $m_\chi \ge m_Y$};\\
$\bullet$~Yellow:
rapid annihilation directly into SM particles via the $s$-channel $Y$ resonance, {in the region where $0.9 <m_Y/(2m_\chi) < 1.1$}. \\}
We see clearly two separated
favoured regions, one with $m_\chi \simeq m_Y/2$ where rapid annihilation via the $Y$ funnel dominates,
and another with $m_\chi > m_Y$ where annihilation into pairs of mediator particles $Y$ dominates.
The boundaries of both allowed regions are
very sharp, reflecting the steepness of the resonance peak in the $s$-channel case and the
phase-space limit in the $t$-channel case~\footnote{{In other projections, the $s$- and $t$-channel regions may overlap, and have quite similar $\chi^2$, so that points with different colours appear interspersed.}}.

We see that the $t$-channel region is located at smaller $m_Y$ than the $s$-channel region for any value of $m_\chi$,
subject to the kinematic constraint $m_Y < m_\chi$.
In this region the annihilation cross-section of this channel, $\chi \chi \to Y Y$,
is proportional to $g_{\rm DM}^4$ and independent of $g_{\rm SM}$.
Therefore, the relic abundance can be brought to the observed value by tuning $g_{\rm DM}$,
unless $m_\chi \lesssim {100}$ GeV,
whilst the experimental constraints from the LHC and the direct DM detection can be avoided
by taking $g_{\rm SM}$ small enough.}

\begin{figure}[htb!]
\centering
\includegraphics[width=0.475\textwidth]{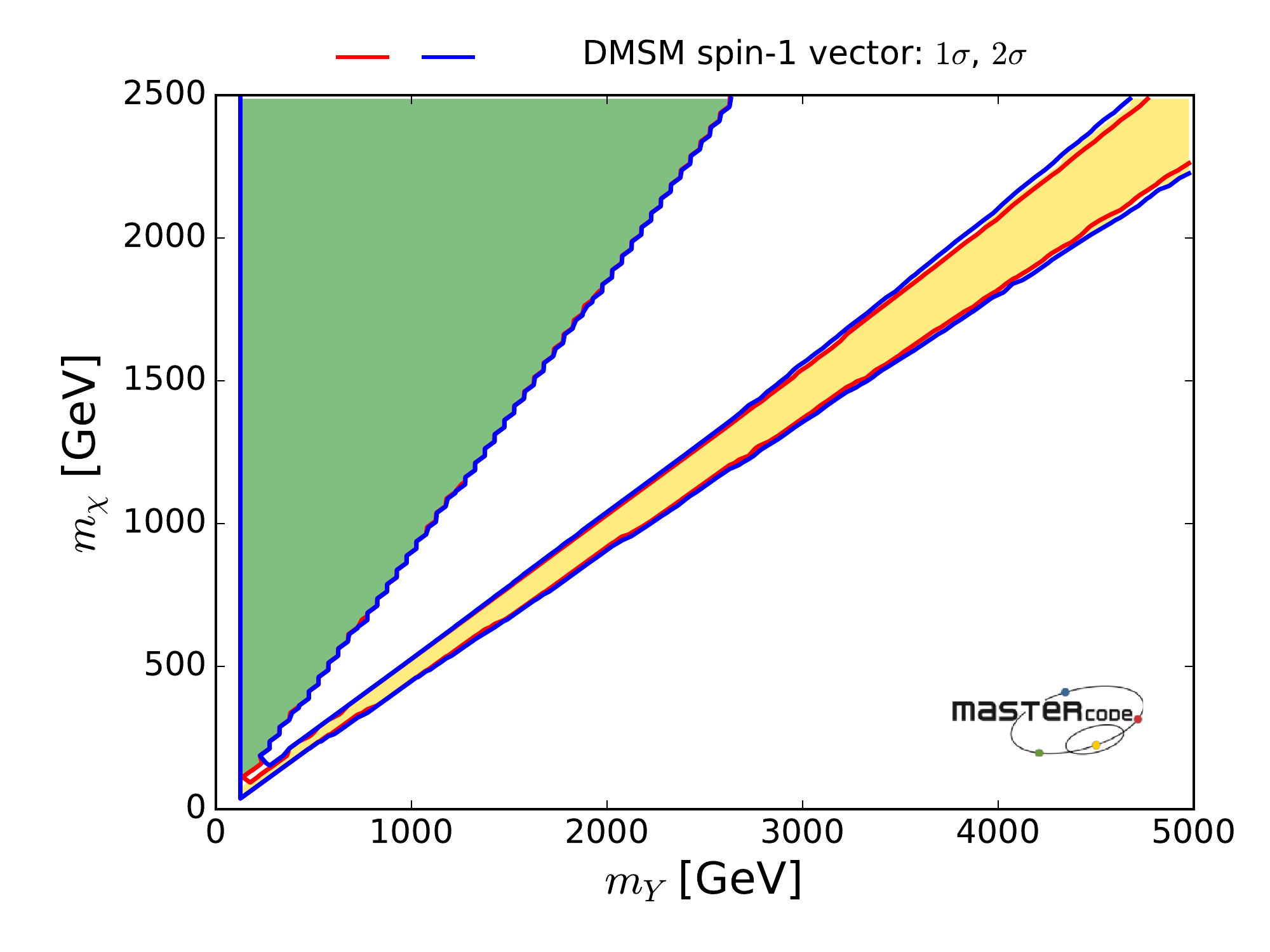}
\caption{\it Preferred regions in the $(m_Y, m_\chi)$ plane in the vector-like model.
We delineate with red (blue) contours, respectively, the parameter regions
with $\Delta \chi^2 < 2.30 (5.99)$, which are favoured at the 68\% (95\%)~CL
and regarded as proxies for 1- (2-)$\sigma$ regions, respectively. We also use colour coding to illustrate the
dominant mechanisms bringing the DM density into the allowed range: green for annihilation via $t$-channel
exchange into pairs of mediator particles $Y$ that subsequently decay into SM particles, and yellow for
rapid annihilation directly into SM particles via the $s$-channel $Y$ resonance.}
\label{fig:vectormassplane}
\end{figure}

{As can be seen in Fig.~\ref{fig:vectormassplane}, the lower limit $m_Y > 100 \gev$ in our sample
enforces a corresponding lower limit {$m_\chi \gtrsim 100 \gev$ in the $t$-channel region. This is because the annihilation process $\chi \chi \to YY$
is kinematically blocked for $m_\chi < m_Y$, so the cross-section for $\chi \chi$ annihilation becomes very small, resulting in DM overdensity. On the other hand, in the $s$-channel region the relic density constraint requires \gdm\ to be very small for $m_Y \sim 100 \gev$, in which case $\Gamma_Y/m_Y \ll 1$ and $\chi \chi$ annihilations are rapid enough only if $m_\chi \simeq m_Y/2 \gtrsim 50 \gev$}, {the inequality being due to our scanning limit on $m_Y$.
The uncoloured vertical band at small mass in this and subsequent figures is a reflection of this restriction
{on the range of $m_Y$ sampled}.}

{We find that the $\Delta \chi^2$ function is negligible
for all values of $m_Y$ above the 100~GeV cut, and we also find that the $\Delta \chi^2$ function is also very small and essentially featureless
for $m_\chi > 50 \gev$.} {Moreover, we find no upper limits on $m_{\chi,Y}$ within the ranges sampled. Our numerical results indicate that
\begin{equation}
\Omega_\chi h^2 \sim 0.1 \left( \frac{m_Y}{2\,{\rm TeV}} \right)^2 \left( \frac{\gdm}{0.01} \right)^{-2}    
\label{OmegaApprox}
\end{equation}
for intermediate values of \gsm, with larger values of $\Omega_\chi h^2$ when \gsm\ is either large or small. Extrapolation of this approximation suggests that the upper bound on $m_Y$ in the $s$-channel region is $m_Y \sim \mathcal{O}(700) \tev$ for $\gdm \lesssim \sqrt{4 \pi}$. 
}

Fig.~\ref{fig:vectorlogcouplingplane} uses logarithmic scales to show the $(\gsm, \gdm)$ plane in the vector-like
model. Again, we distinguish immediately two distinct preferred regions,
{separated where $\gsm \lesssim 3 \times 10^{-4}$
and $\gdm \gtrsim 0.3$.}
As can be seen from the colour coding,
the {region at small \gsm\ and large \gdm\ corresponds} to the triangular $t$-channel region
in Fig.~\ref{fig:vectormassplane}, whereas the larger values of {\gsm\, and \gdm\ outside this region}
appear in the  $s$-channel funnel region.
{This region is bounded at large values of the product \gsm \gdm\ by the upper limit on \ssi\ and the limited scanning
range $m_Y < 5 \tev$, as indicated [see Eq.~(\ref{eq:ssi_expression})]: the upper right portion of the figure with larger values of \gsm \gdm\ would be allowed for larger $m_Y$. 
In the region $0.07 > \gdm > 3 \times 10^{-4}$, $\Omega h^2 \propto m_Y^2 / \gdm^2$ and can be sufficiently small
if $m_\chi \simeq m_Y/2$ and $m_Y < 5 \tev$.
In this case, Eqs.~\eqref{eq:ssi_expression} and \eqref{OmegaApprox} imply
\be
\gsm \lesssim 0.1 \cdot \left( \frac{m_\chi}{1\,{\rm TeV}} \right)
\left[ \frac{\sigma_{\rm SI}^{\rm UL}(m_\chi)}{10^{-45} [{\rm cm}^{-2}]} \right]^{\frac{1}{2}}
\label{eq:gsmlim}
\ee
where $\sigma_{\rm SI}^{\rm UL}(m_\chi)$ is the mass-dependent experimental upper bound on $\ssi$. We note also that the dijet constraint is the strongest at 
the largest value of $\gsm \sim 0.3$. Thus the lower right part of the figure is excluded by the upper limit on the elastic scattering cross section.
Finally, the apparent exclusion in the lower region in the figure, where $\gdm \lesssim 3 \times 10^{-4}$, {rising at smaller \gsm,} is also an artefact of the scanning limit $m_Y > 100 \gev$
{combined with the relic density constraint},
as indicated.}

\begin{figure}[htb!]
\centering
\includegraphics[width=0.475\textwidth]{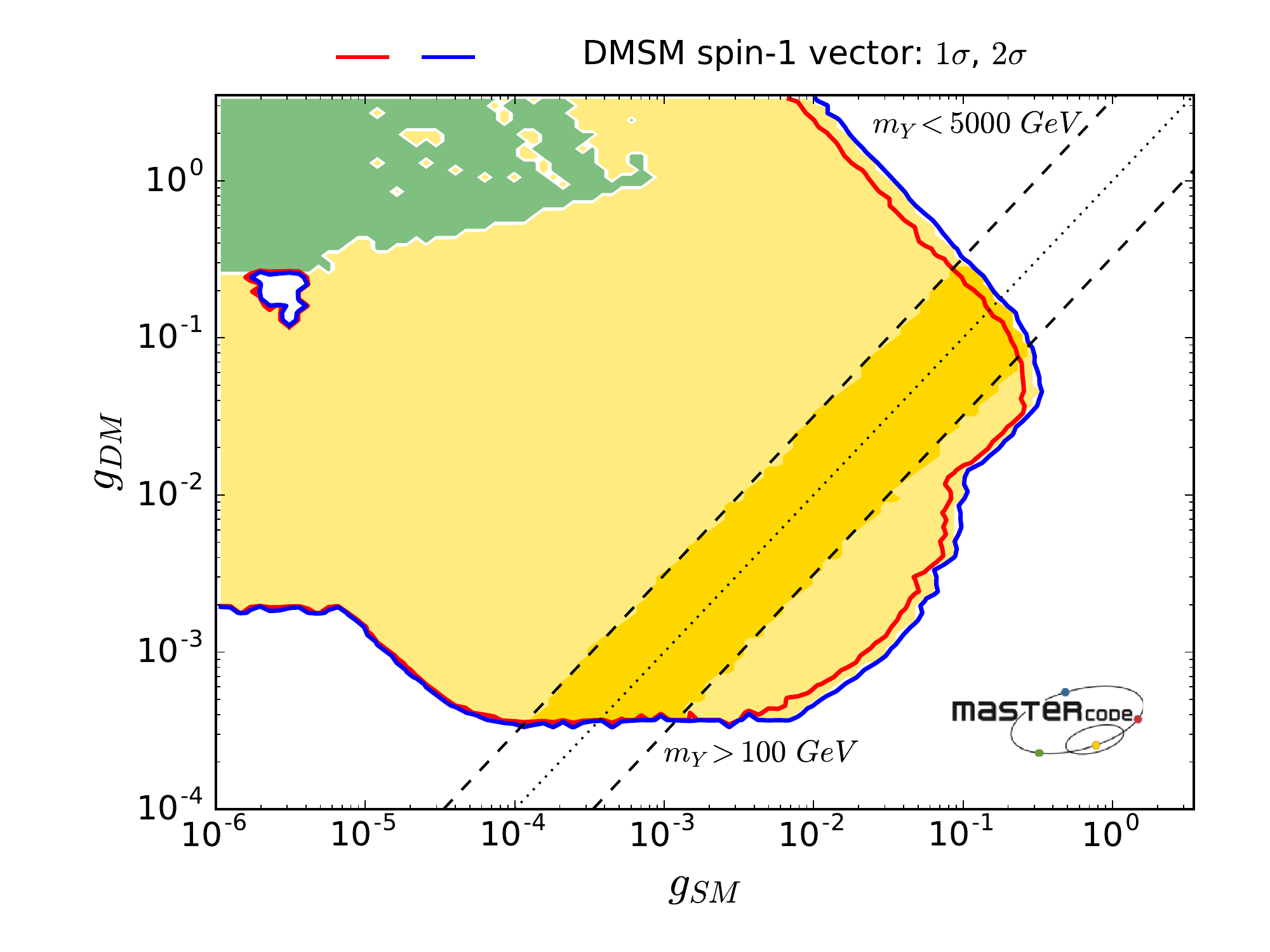}
\caption{\it Preferred regions in the $(\gsm, \gdm)$ plane (on logarithmic scales) in the vector-like model, 
where we have marginalized over the masses, $m_\chi$ and $m_Y$.
We again use colour coding to illustrate the
dominant mechanisms bringing the DM density into the allowed range: green for annihilation via $t$-channel
exchange into pairs of mediator particles $Y$ that subsequently decay into SM particles, and yellow for
rapid annihilation directly into SM particles via the $s$-channel $Y$ resonance. {The diagonal dotted line
indicates where $\gdm = \gsm$, and the band where $1/3 < \gdm/gsm < 3$ is bounded by dashed lines and shaded a darker yellow.}}
\label{fig:vectorlogcouplingplane}
\end{figure}

We have indicated by a diagonal dotted line where $\gdm = \gsm$. As discussed later, we might expect
$\gdm$ and $\gsm$ to be similar in magnitude in many UV completions of DMSMs, and we study later the
restriction to {the band between the dashed lines in Fig.~\ref{fig:vectorlogcouplingplane} 
that is shaded darker yellow}, where $1/3 < \gdm/\gsm < 3${, see Section~\ref{sec:gdmeqgsm}.}

{We do not show the one-dimensional likelihood function for \gsm, which is quite
featureless {apart from a sharp rise for $\gsm \gtrsim 0.3$}, nor that for \gdm, which is also featureless
apart from a steep rise for $\gdm \lsim 3 \times 10^{-4}$ that
is an artefact of the limit on our scanning range for $m_Y$.}

{Fig.~\ref{fig:vectorcouplingmass} displays the planes of $m_Y$ and the two couplings $\gsm,\ \gdm$. We see in the left panel that
within the $t$-channel region there is a strong upper limit on $\gsm \lesssim 10^{-3}$,}
which is enforced by the combination of 
the upper limit on \ssi, which constrains the product $\gdm \, \gsm$,
and the lower limit on $\gdm$
{visible in the right panel, which is another result of our limited scan to $m_Y > 100 \gev$}.
{Since $\ssi$ scales as $\gdm^2 \gsm^2/m_Y^4$, the limit is stronger for smaller $m_Y$, 
and $\gsm$ must be smaller than $10^{-4}$ for $m_Y \lesssim 100$ GeV.}
The upper limit on $m_Y$ in the $t$-channel region
visible for small \gsm\ is an artefact of the limited scanning range {$m_\chi < 2.5 \tev$}.
The upper limit on $\gsm$ in the $s$-channel region
comes {mainly from the LHC dijet constraint, with the \ssi\ constraint
also playing a role when $m_Y \lesssim 500 \gev$.}

\begin{figure*}[htb!]
\centering
\mbox{}\vspace{1em}
\includegraphics[width=0.475\textwidth]{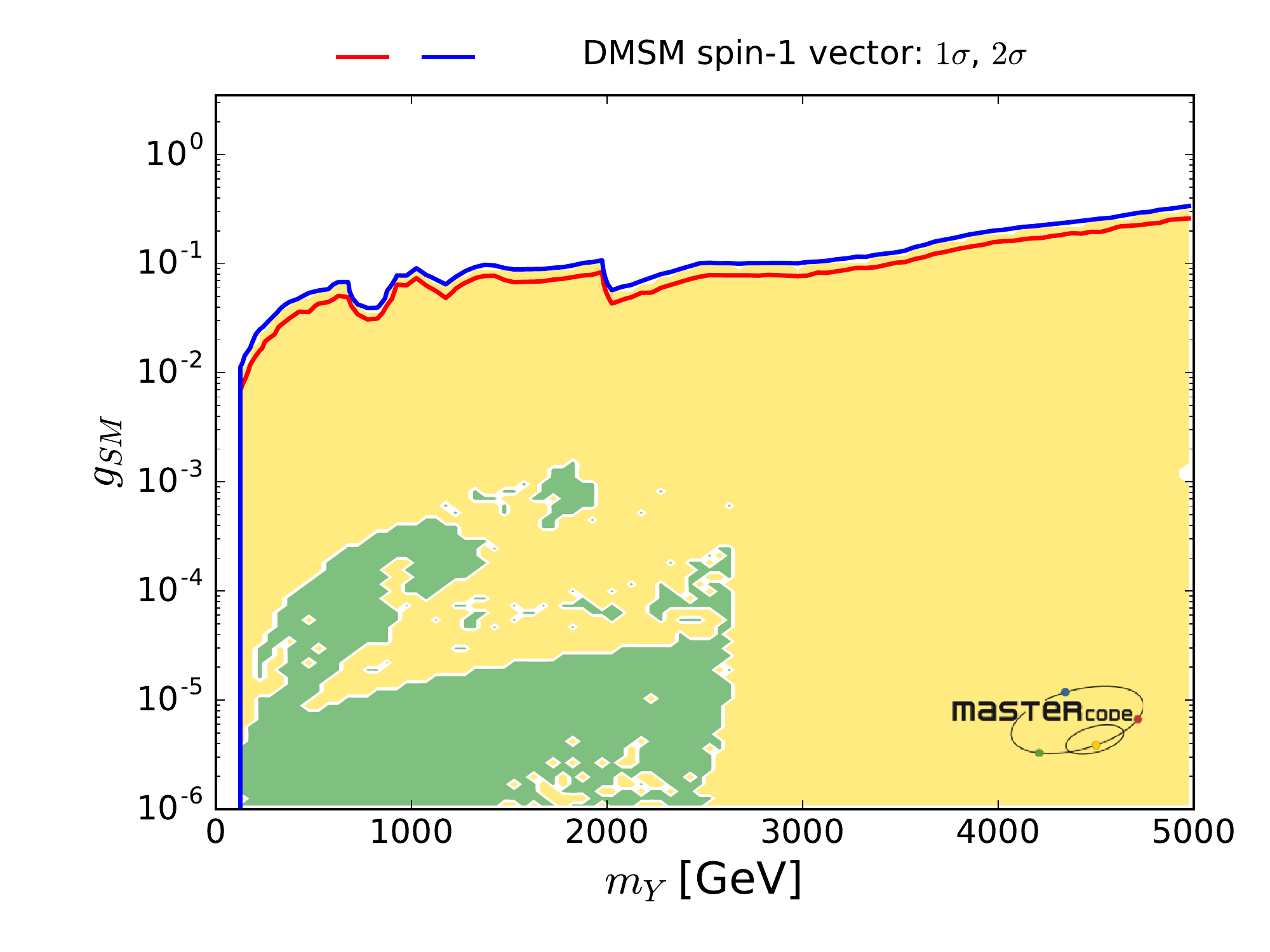}
\includegraphics[width=0.475\textwidth]{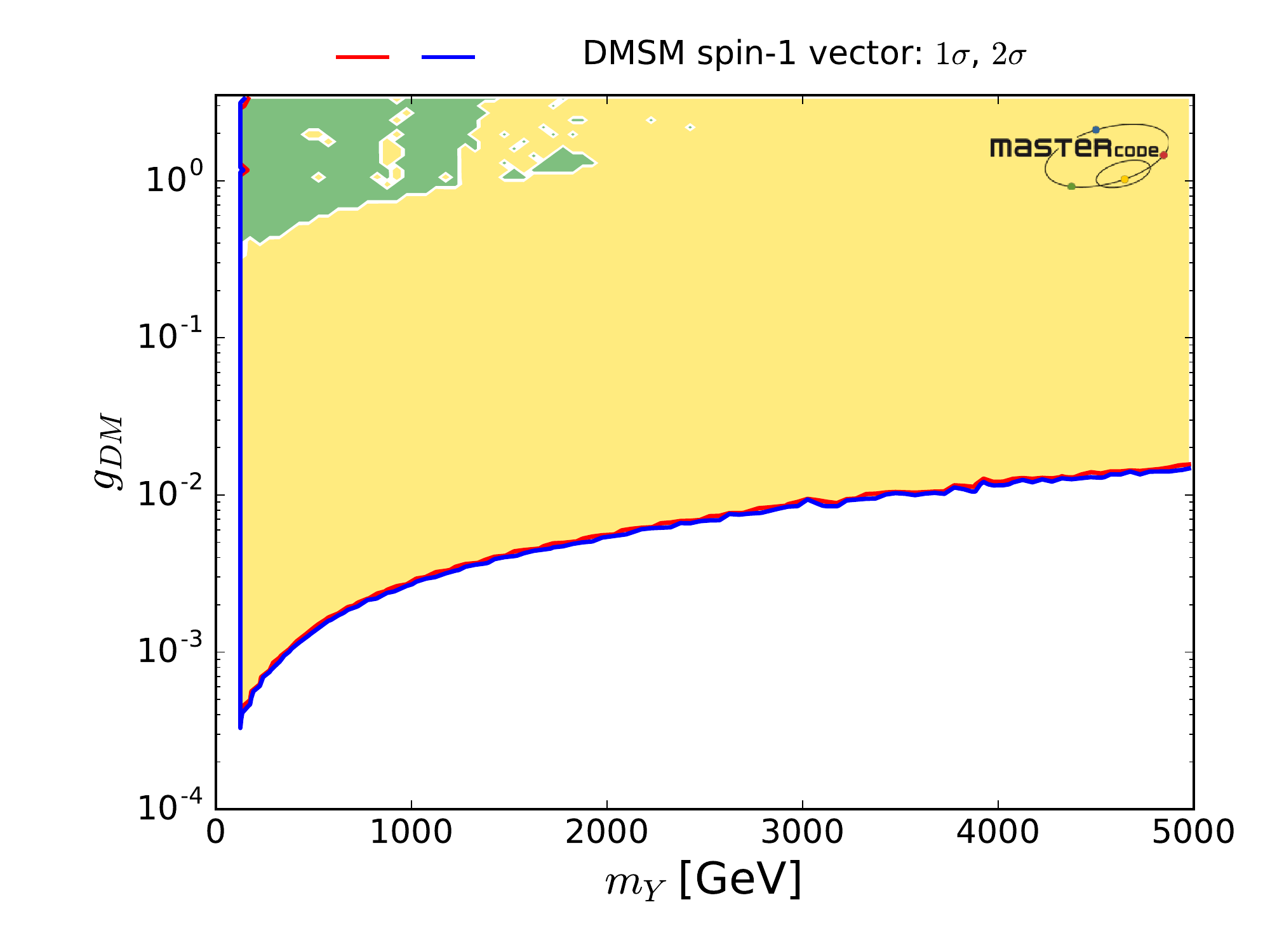}
\caption{\it The likelihood functions for $\gsm$ (left panel) and $\gdm$ (right panel)
as functions of $m_Y$ in the vector-like model. We again use colour coding to illustrate the
dominant mechanisms bringing the DM density into the allowed range: green for annihilation via $t$-channel
$\chi$ exchange into pairs of mediator particles $Y$ that subsequently decay into SM particles, and yellow for
rapid annihilation directly into SM particles via the $s$-channel $Y$ resonance.}
\label{fig:vectorcouplingmass}
\end{figure*}

{We see in the left panel of Fig.~\ref{fig:vectorcouplingmass}
that $\gsm \lesssim 10^{-2}$ for $m_Y \sim 100 \gev$ and
$\lesssim 0.1$ for $m_Y > 1 \tev$. Comparing with Fig.~11 of~\cite{Ellis:2018xal}, where leptophobic models with
universal quark couplings were analyzed~\footnote{{We recall that in such models the $Z$ and $Y$ have only kinetic mixing,
which is loop-induced and hence
suppressed.}}, and identifying \gsm\ with the combination $g Y^\prime_q$
of the parameters defined in those models, 
we see that the precision electroweak data do not constrain our model sample.}

In the right panel of Fig.~\ref{fig:vectorcouplingmass}, the lower bound on \gdm\ in the $t$-channel region
at low $m_Y$ is due to the DM density constraint. 
{In the $s$-channel region the relic constraint imposes a weaker bound on \gdm,
which is given roughly by
\begin{equation}
\Big( \frac{m_Y}{2 \,\rm TeV} \Big)^2 \Big( \frac{\gdm}{0.01} \Big)^{-2} \lesssim 1 \,,
\label{rough}
\end{equation}
}
{as follows from Eq.~\eqref{OmegaApprox} for $2 m_{\chi} \sim m_Y$.}

\begin{figure*}[htb!]
\centering
\vspace{3em}
\includegraphics[width=0.475\textwidth]{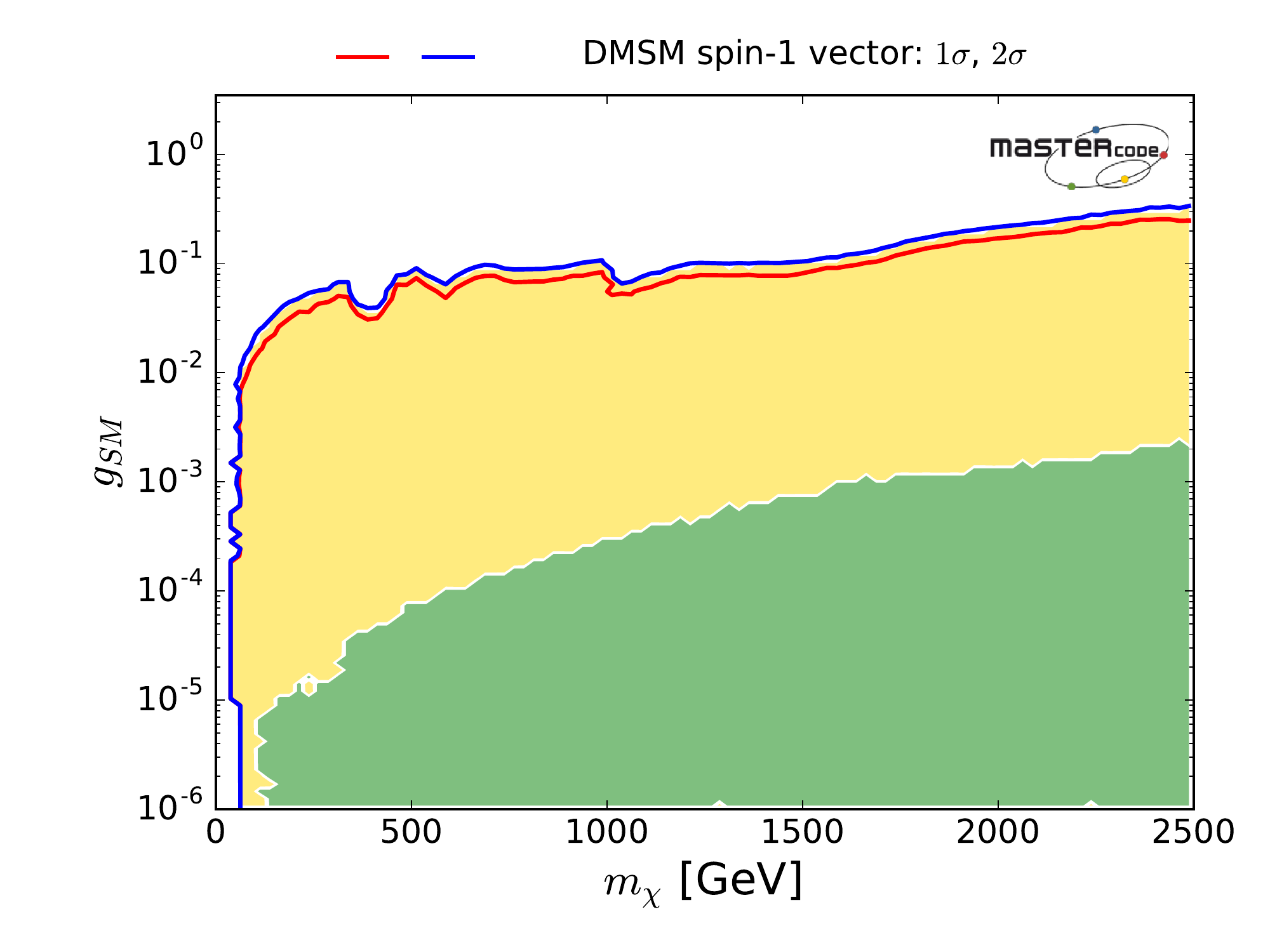}
\includegraphics[width=0.475\textwidth]{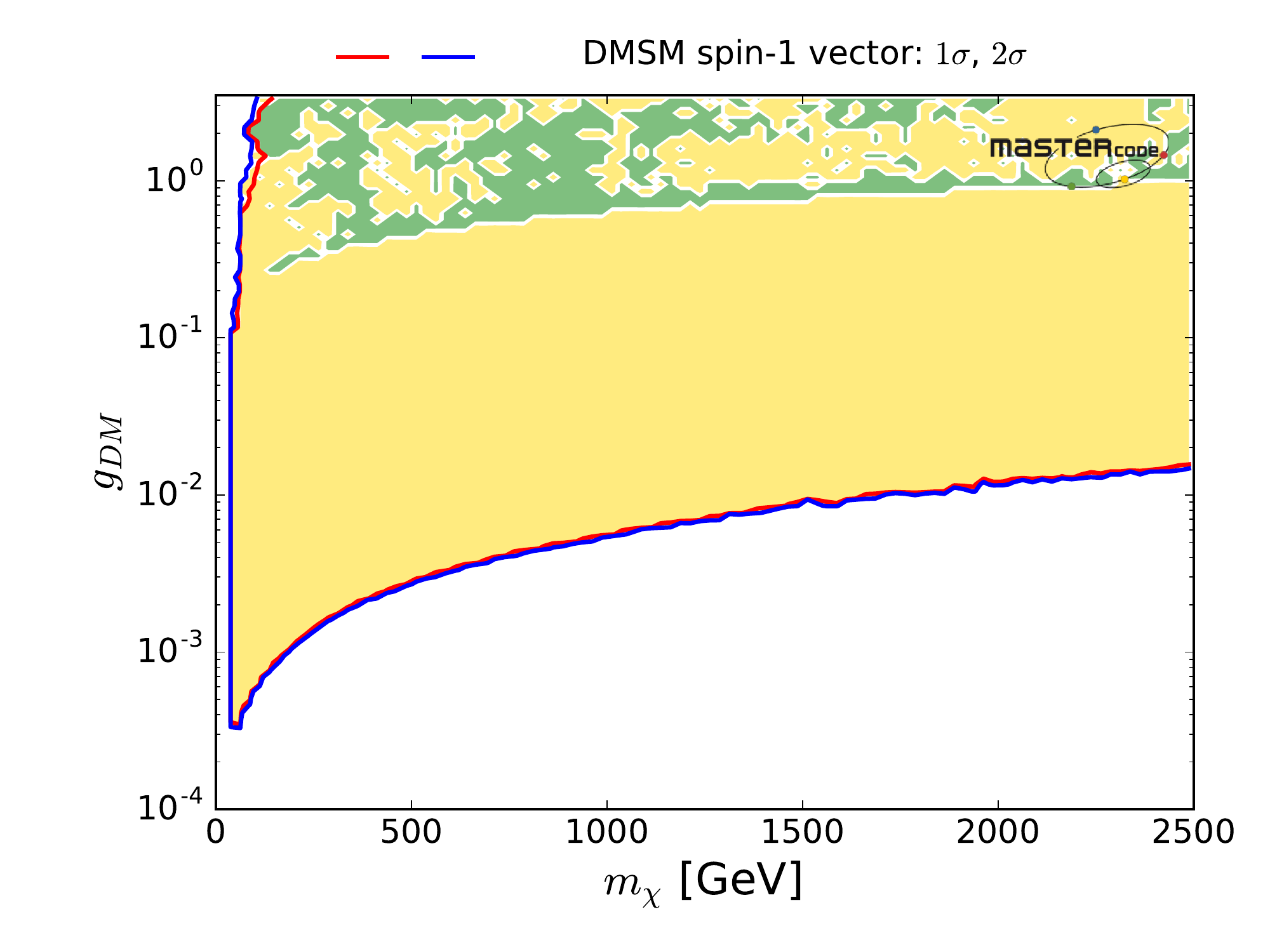}
\caption{\it The likelihood functions for $\gsm$ (left panel) and $\gdm$ (right panel)
as functions of $m_\chi$ in the vector-like model. We again use colour coding to illustrate the
dominant mechanisms bringing the DM density into the allowed range: green for annihilation via $t$-channel
$\chi$ exchange into pairs of mediator particles $Y$ that subsequently decay into SM particles, and yellow for
rapid annihilation directly into SM particles via the $s$-channel $Y$ resonance.}
\label{fig:DMvectormass}
\end{figure*}

{Fig.~\ref{fig:DMvectormass} displays the $(m_\chi, \gsm)$ plane in the
left panel and the $(m_\chi, \gdm)$ plane in the right panel. 
The upper bounds on \gsm\ and the lower bounds on \gdm\ are the same as in Fig.~\ref{fig:vectorcouplingmass},
all increasing with $m_\chi$}. 

Fig.~\ref{fig:ssi} shows the likelihood function in the $(m_\chi, \ssi)$ plane for the vector-like model.
We see that, as already visible in the left panel of Fig.~\ref{fig:DMvectormass}, {only values of $m_\chi \gtrsim 50$~GeV
are allowed, {for the reason mentioned previously, namely the interplay of the cut $m_Y > 100 \gev$ (as indicated) and the relic density constraint.} 
Above this value, a large range of values of \ssi\ is allowed in both the $t$- and $s$-channel regions.
The upper limits on \ssi\ at various
confidence levels are determined by the combined experimental likelihood for the LUX~\cite{LUX}, PANDAX-II~\cite{PANDAX} and
XENON1T~\cite{XENON} experiments, which we have rescaled to account for the different local DM
density that we assume. In the $s$-channel region, small values of \ssi\
are allowed when $m_\chi \sim m_Y/2$ and small values of \gdm\ and/or \gsm\ are favoured by
the relic density constraint, and small values of \ssi\ are allowed in the $t$-channel region
because small values of \gsm\ are allowed, as discussed previously. On the other hand,
we see that \ssi\
may well lie within the range to be probed by upcoming experiments such as LUX-ZEPLIN (LZ)~\cite{LZ} and XENONnT~\cite{XENONnT}, though \ssi\ may also be much smaller than the current experimental
sensitivity, even below the neutrino `floor' {indicated by the dashed orange line in Fig.~\ref{fig:ssi}, which is based on~\cite{McCabe}, updating~\cite{floor}}.}

\begin{figure}[htb!]
\centering
\includegraphics[width=0.475\textwidth]{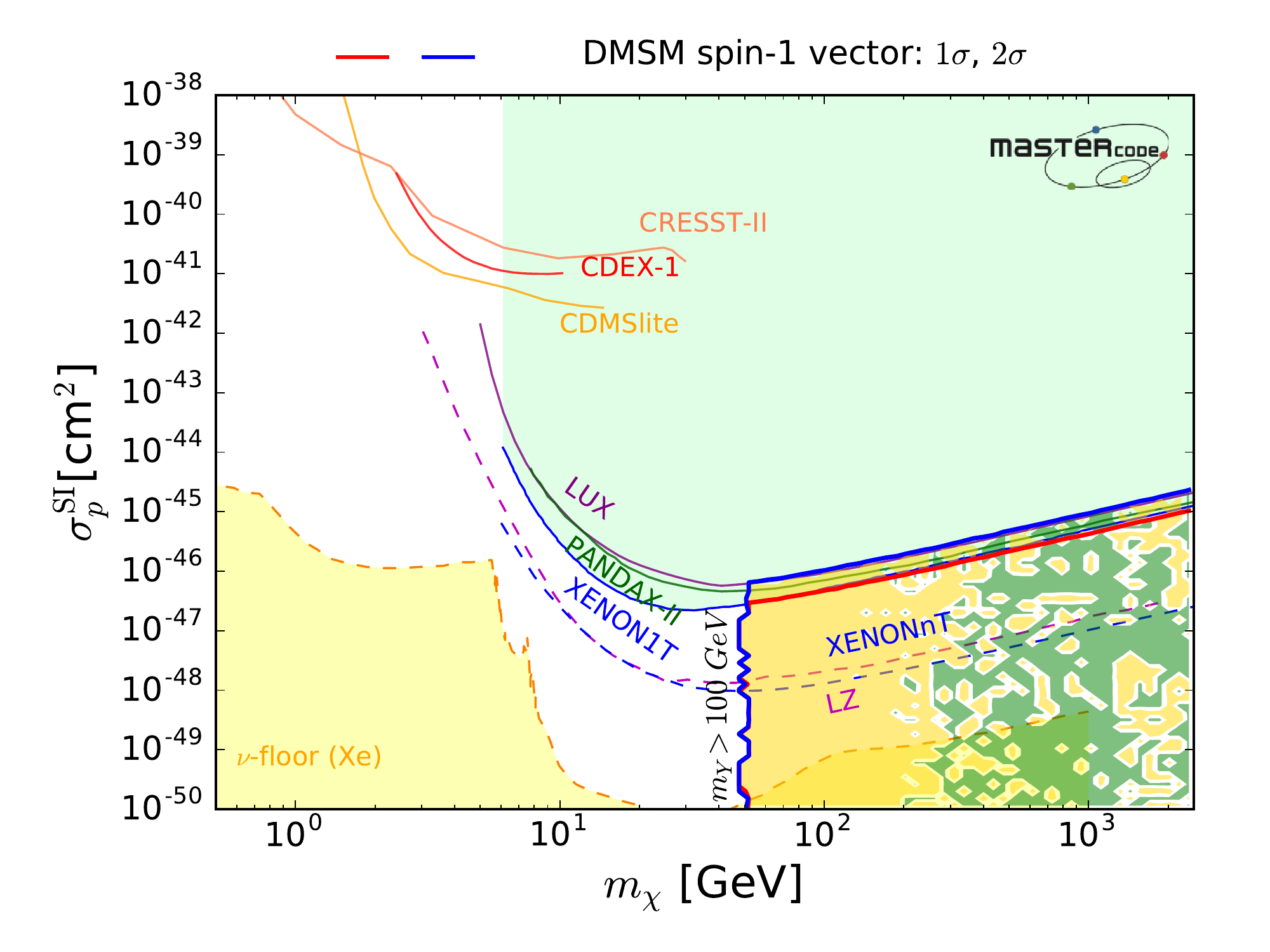}
\caption{\it Contours of the likelihood in the $(m_\chi, \ssi)$ plane for the vector-like model, showing
the current upper limits from the LUX~\protect\cite{LUX}, PANDAX-II~\protect\cite{PANDAX} and
XENON1T~\protect\cite{XENON} experiments (rescaled to account for the different local DM
density that we assume), together with the neutrino `floor'~\protect\cite{floor} (shown as
the dashed orange line),
and the range of \ssi\ that will be probed by the upcoming experiments LZ~\protect\cite{LZ} and XENONnT~\protect\cite{XENONnT}.
We again use colour coding to illustrate the
dominant mechanisms bringing the DM density into the allowed range: green for annihilation via $t$-channel
$\chi$ exchange into pairs of mediator particles $Y$ that subsequently decay into SM particles, and yellow for
rapid annihilation directly into SM particles via the $s$-channel $Y$ resonance.
{We indicate the effective lower limit on $m_\chi$ that is imposed by our sampling limit on $m_Y$.}}
\label{fig:ssi}
\end{figure}


\subsection{Axial-Vector DM Couplings}

We now turn to the case of DM with axial-vector couplings. Fig.~\ref{fig:axialvectormassplane} displays the
$(m_Y, m_\chi)$ plane in this case: it is also colour-coded according to the dominant DM annihilation
mechanism using the same shading scheme as for the vector case, and
the 1- and 2-$\sigma$ contours are indicated by red and blue lines, respectively. 
We see an $s$-channel funnel
feature that is {rather} broader than in the case of DM with vector couplings shown in Fig.~\ref{fig:vectormassplane}, {and in this case we shade yellow the region where $0.6 < m_Y/(2 m_\chi) < 2$}.

{This {broadening} occurs because, whereas the direct detection constraint is very severe for the vector mediator,
so that $g_{\rm DM} \, g_{\rm SM}$ cannot be large and the parameters {should be near the peak of the resonance where
$m_Y \sim 2 m_{\chi}$, in order for the DM particles to annihilate} sufficiently (see Fig.~2 of~\cite{Kahlhoefer:2015bea} and the
accompanying text), the strong \ssi\ constraint is absent in the axial-vector case, {so that off-resonance regions of parameter space with larger
values of $\gsm \, \gdm$ are allowed where annihilation is} $p$-wave or {$m_q$} suppressed (see the discussion around Eq.~(7) of~\cite{1403.4837}). 
This opens up more parameter space in the $(m_{Y}, m_{\chi})$ plane, with the
deviation from $m_Y \sim 2 m_{\chi}$ bounded only by the dijet constraint. We note that this constraint is weaker
when $m_Y > 2 m_\chi$ because the decay channel $Y \to \chi \chi$ is open.
{We also note that, as in the vector case, at low masses
the funnel region merges with the $t$-channel annihilation region 
where $m_\chi > m_Y$, so that the
preferred parameter space is simply connected also in the axial-vector case.}

\begin{figure}[htb!]
\centering
\includegraphics[width=0.475\textwidth]{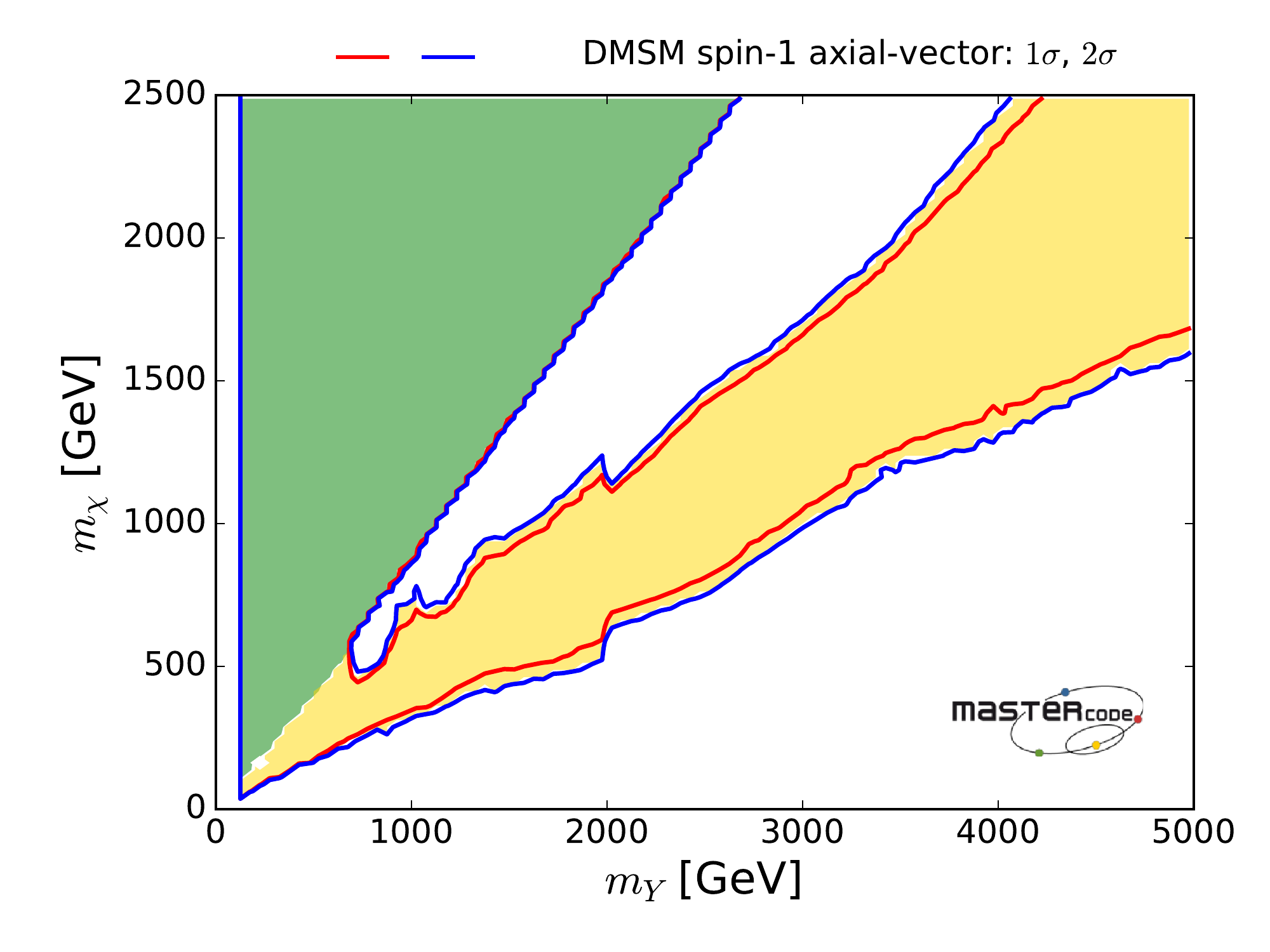}
\caption{\it Preferred regions in the $(m_Y, m_\chi)$ plane in the model with axial-vector DM couplings.
We delineate with red (blue) contours, respectively, the parameter regions
with $\Delta \chi^2 < 2.30 (5.99)$, which are favoured at the 68\% (95\%) CL
and regarded as proxies for 1- (2-)$\sigma$ regions, respectively. We use colour coding to illustrate the
dominant mechanisms bringing the DM density into the allowed range: green for annihilation via $t$-channel
exchange into pairs of mediator particles $Y$ that subsequently decay into SM particles, and yellow for
rapid annihilation directly into SM particles via the $s$-channel $Y$ resonance.}
\label{fig:axialvectormassplane}
\end{figure}

{Also as in the vector-like case, we see again in the axial case in Fig.~\ref{fig:axialvectormassplane}
the lower bound $m_\chi \gtrsim 50 \gev$} {due to the interplay of the sampling limit $m_Y > 100 \gev$ and the relic density constraint. Above this value of $m_\chi$, 
the one-dimensional $\Delta \chi^2$ likelihood function for $m_\chi$ is featureless, as is that for $m_Y > 100 \gev$.}

{In Fig.~\ref{fig:axialvectorlogcouplingplane} we show the $(\gsm, \gdm)$ plane in the axial-vector
model, using logarithmic scales. The $s$-channel region extends to larger values of $\gsm$ and $\gdm$
than in the vector case, {because of the suppression of axial-vector-mediated $s$-channel annihilation discussed above.
 Values of 
$\gdm$ as large as the sampling limit $\sqrt{4 \pi}$ are allowed.}
{We see a separation
between the regions of this plane where the $s$- and $t$-channel mechanisms are dominant when $\gsm \sim 10^{-2}$ {and $\gdm \sim 10^{-1}$}. The upper bound on $\gsm$ comes from the dijet constraint.}
As in the vector case, the dotted line is where $\gsm = \gdm$,
and the deeper shading indicates where $1/3 < \gsm/\gdm < 3${, favouring the {large-\gsm\ part of the}
$s$-channel annihilation region, see the discussion in 
Section~\ref{sec:gdmeqgsm}.} {As in the vector case, there is a
region at low \gdm, rising at small \gsm, whose exclusion by the relic
density constraint is an artefact of the
sampling restriction $m_Y > 100 \gev$.}

\begin{figure}[htb!]
\centering
\includegraphics[width=0.475\textwidth]{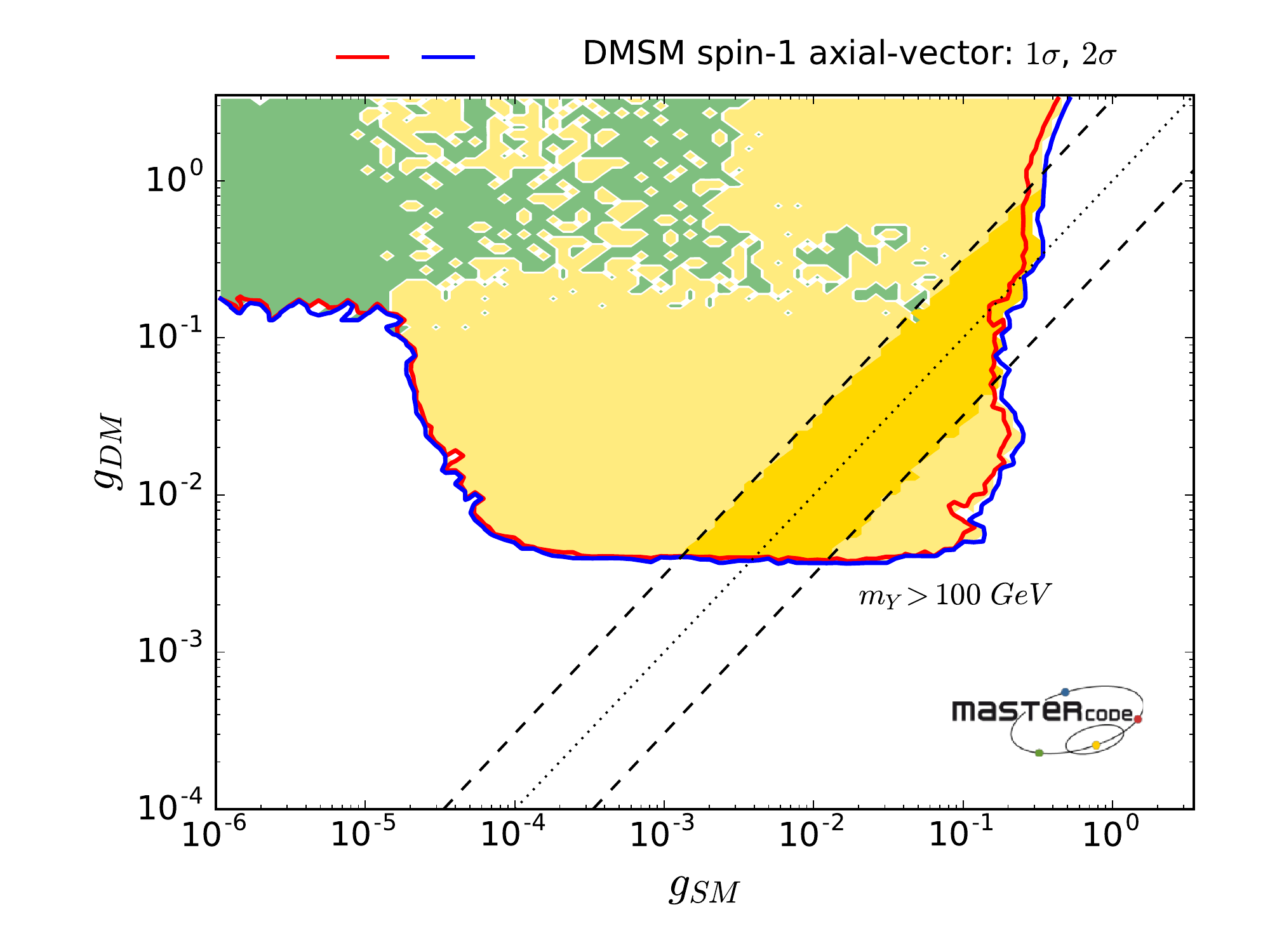}
\caption{\it Preferred regions in the $(\gsm, \gdm)$ plane (on logarithmic scales) in the model with axial-vector DM couplings,
again using colour coding to illustrate the
dominant mechanisms bringing the DM density into the allowed range: green for annihilation via $t$-channel
exchange into pairs of mediator particles $Y$ that subsequently decay into SM particles, and yellow for
rapid annihilation directly into SM particles via the $s$-channel $Y$ resonance. {The diagonal dotted line
indicates where $\gdm = \gsm$, and the band where $1/3 < \gdm/\gsm < 3$ is bounded by dashed lines and shaded a darker yellow.}}
\label{fig:axialvectorlogcouplingplane}
\end{figure}

We display in Fig.~\ref{fig:axialgSMandgDMvsmY} the $(m_Y, \gsm)$ and $(m_Y, \gdm)$ planes (left and right
panels, respectively) in the scenario with axial-vector couplings. We see in the left panel that the $t$-channel DM
mechanism {is important for $m_Y \lesssim 2.6 \tev$. The limit visible in
Fig.~\ref{fig:axialvectormassplane}, which is due to the sampling limit $m_\chi < 2.5 \tev$. Annihilation via the $s$-channel
becomes more important as $m_Y$ increases, as also seen in
Fig.~\ref{fig:axialvectormassplane}, and is the only mechanism for
$m_Y \gtrsim 2.6 \tev$ up to the sampling limit of $5 \tev$.} 
The upper limit on \gsm\ for $m_Y < 5 \tev$ is due to the dijet constraint. 
{Unlike the vector case seen in the left panel of Fig.~\ref{fig:vectorcouplingmass}, the $s$-channel region is excluded for small $\gsm$.  This is due to the $p$-wave nature of the $s$-channel annihilation via the axial mediator, which leads to the scaling behaviour 
$\Omega h^2 \approxprop m_Y^2/(\gsm^2 \gdm^2)$,
as discussed around Eq.~\eqref{Omegagdm_axial}.}

We see again in the right panel that the $t$-channel mechanism is {relevant for $m_Y \lesssim 3 \tev$, whereas the $s$-channel
mechanism is dominant for larger $m_Y$.} 
{The lower limit on $\gdm$ is given by the relic density constraint
since. even exactly at $m_Y = 2 m_\chi$, { $\Omega h^2 \approxprop m_Y^2/(\gsm^2 \gdm^2)$}
and DM is overproduced for sufficiently small \gdm.}

\begin{figure*}[htb!]
\centering
\mbox{}\vspace{1em}
\includegraphics[width=0.475\textwidth]{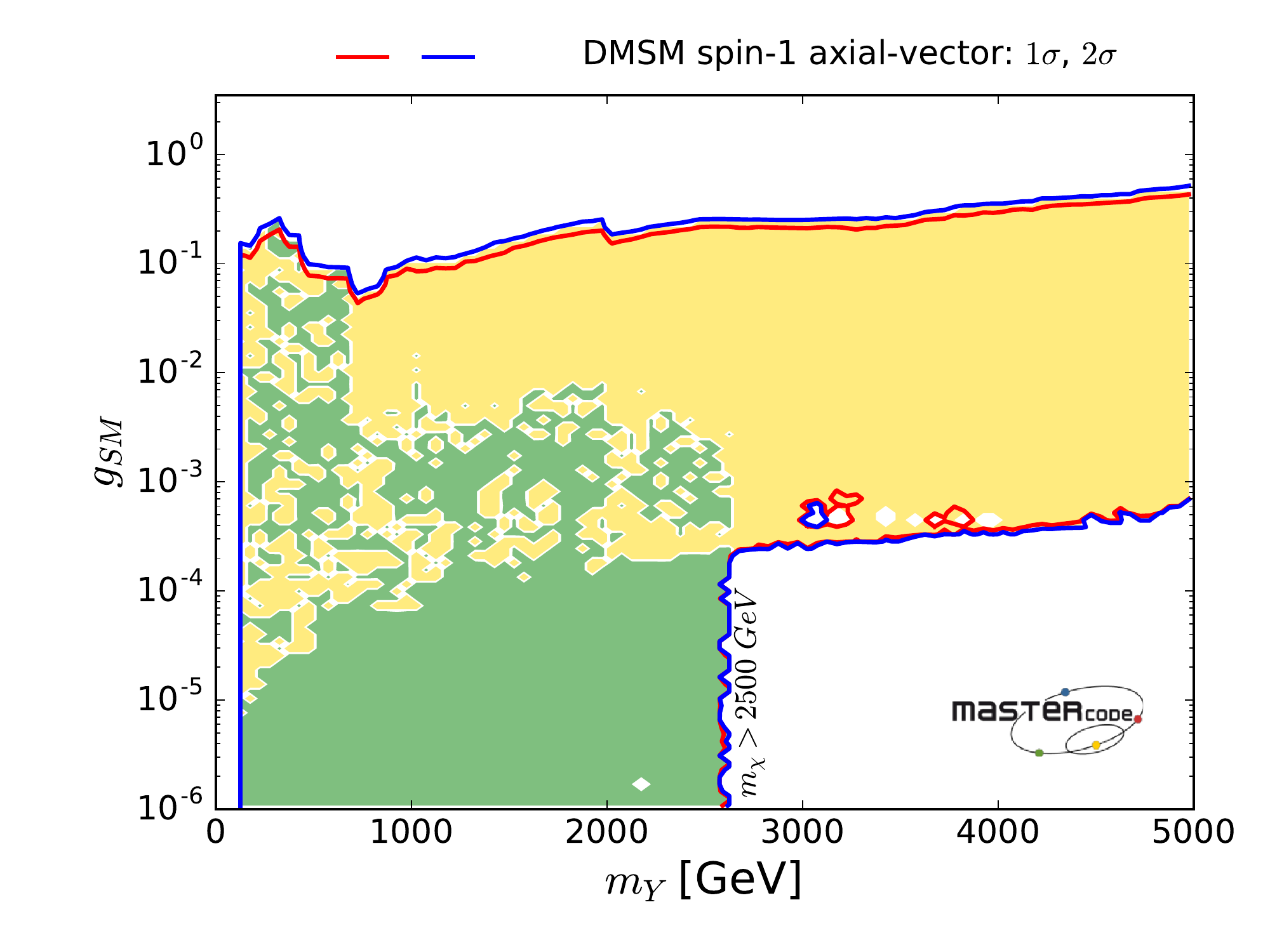}
\includegraphics[width=0.475\textwidth]{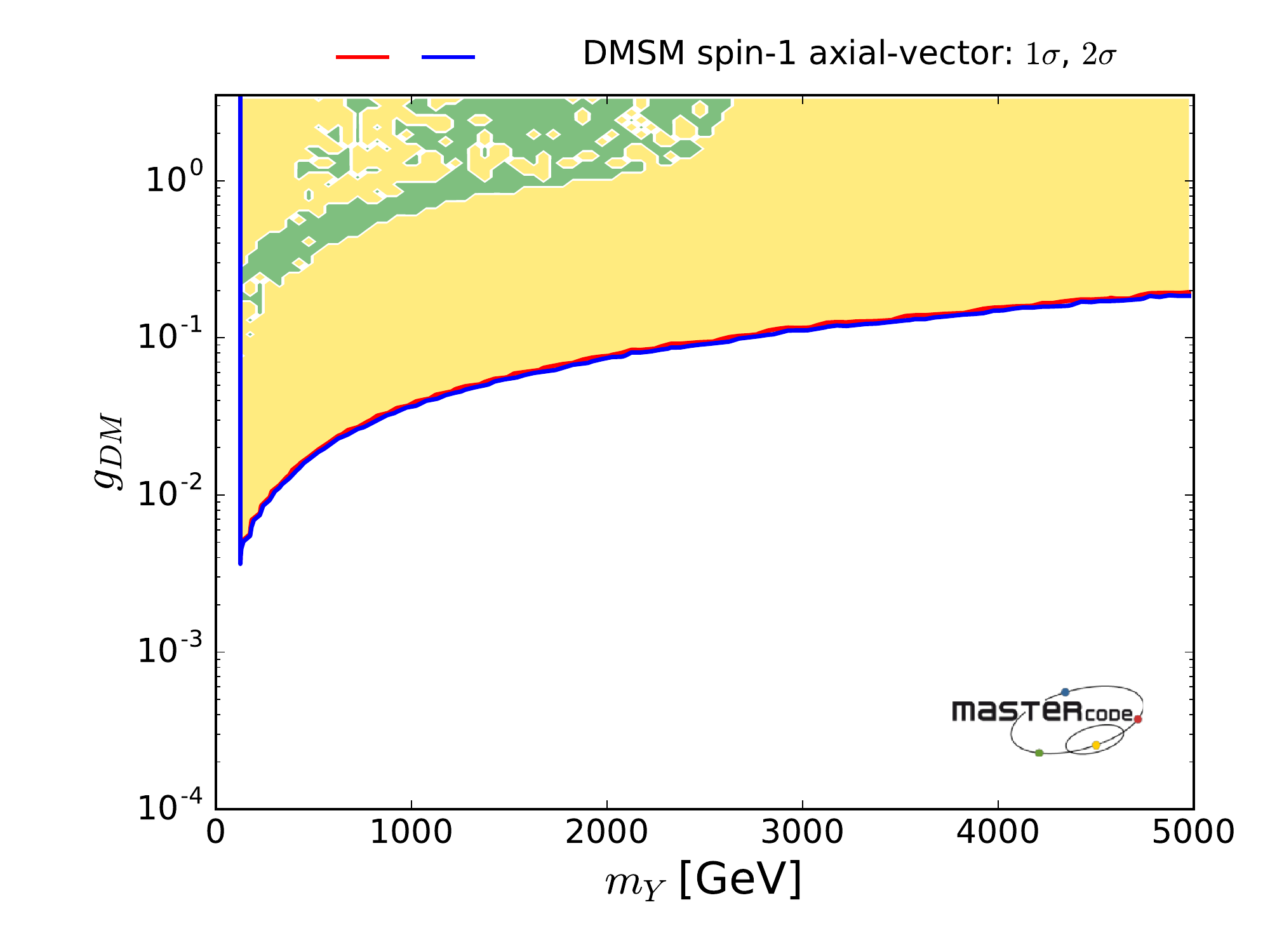}
\caption{\it The likelihood functions for $\gsm$ (left panel) and $\gdm$ (right panel)
as functions of $m_Y$ in the model with axial-vector couplings. We again use colour coding to illustrate the
dominant mechanisms bringing the DM density into the allowed range: green for annihilation via $t$-channel
$\chi$ exchange into pairs of mediator particles $Y$ that subsequently decay into SM particles, and yellow for
rapid annihilation directly into SM particles via the $s$-channel $Y$ resonance.}
\label{fig:axialgSMandgDMvsmY}
\end{figure*}

Fig.~\ref{fig:axialgSMandgDMvsmchi} shows the $(m_\chi, \gsm)$ and $(m_\chi, \gdm)$ planes (left and right
panels, respectively) in the axial-vector model. We see again in the left panel that values of $m_\chi$ up to the
sampling limit of $2.5 \tev$ are allowed {both at smaller $\gsm$ where the $t$-channel mechanism dominates
and at larger $\gsm$} where the $s$-channel mechanism dominates.
{There is a region at low $m_\chi$ where the $s$-channel mechanism dominates.}
The lower limit on $\gdm$ comes from the relic density constraint
as can be seen in Eqs.~\eqref{gsm>gdm}, \eqref{gsm<gdm} and \eqref{eq:Oh2_t}.

\begin{figure*}[htb!]
\centering
\vspace{1em}
\includegraphics[width=0.475\textwidth]{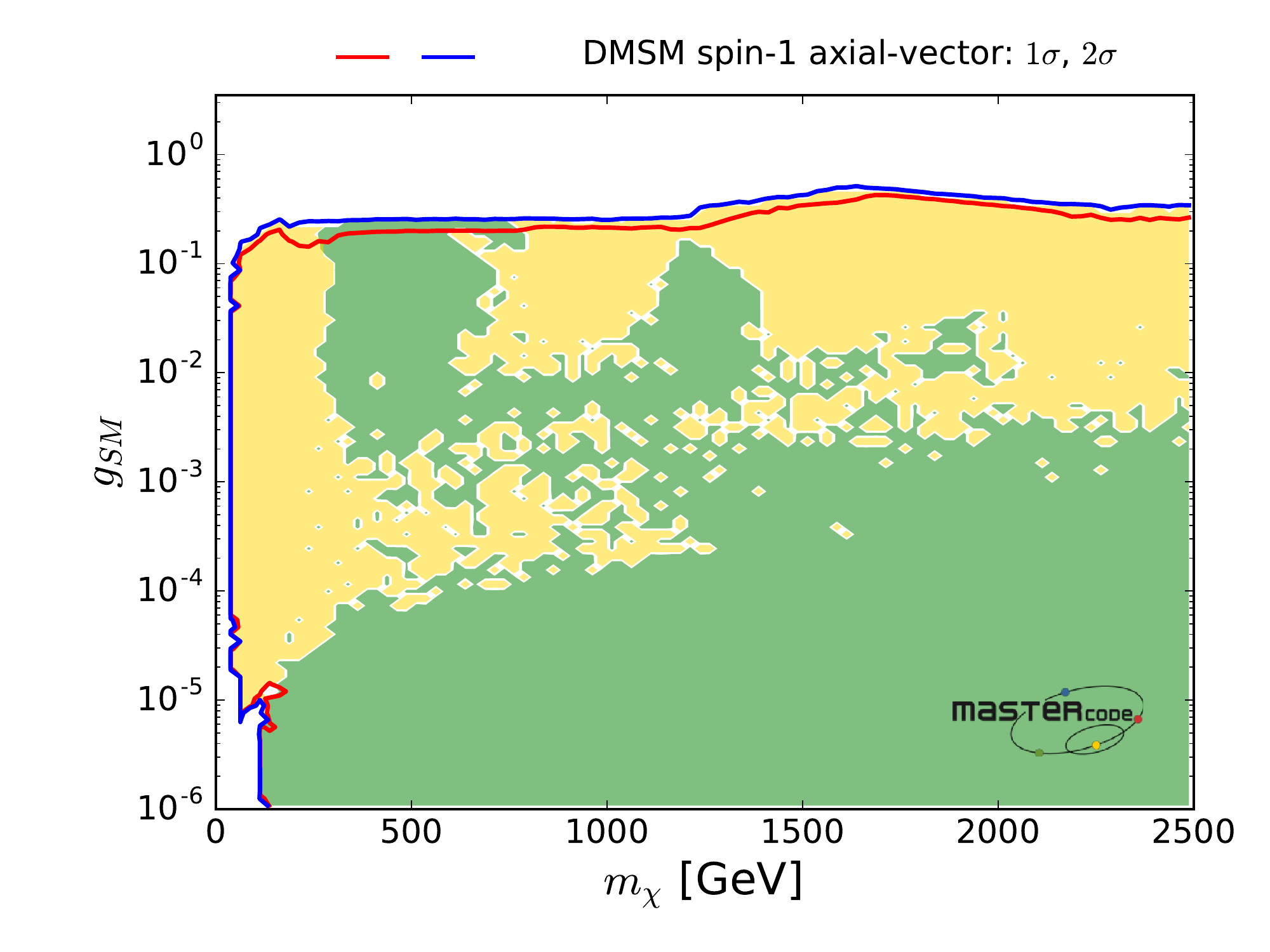}
\includegraphics[width=0.475\textwidth]{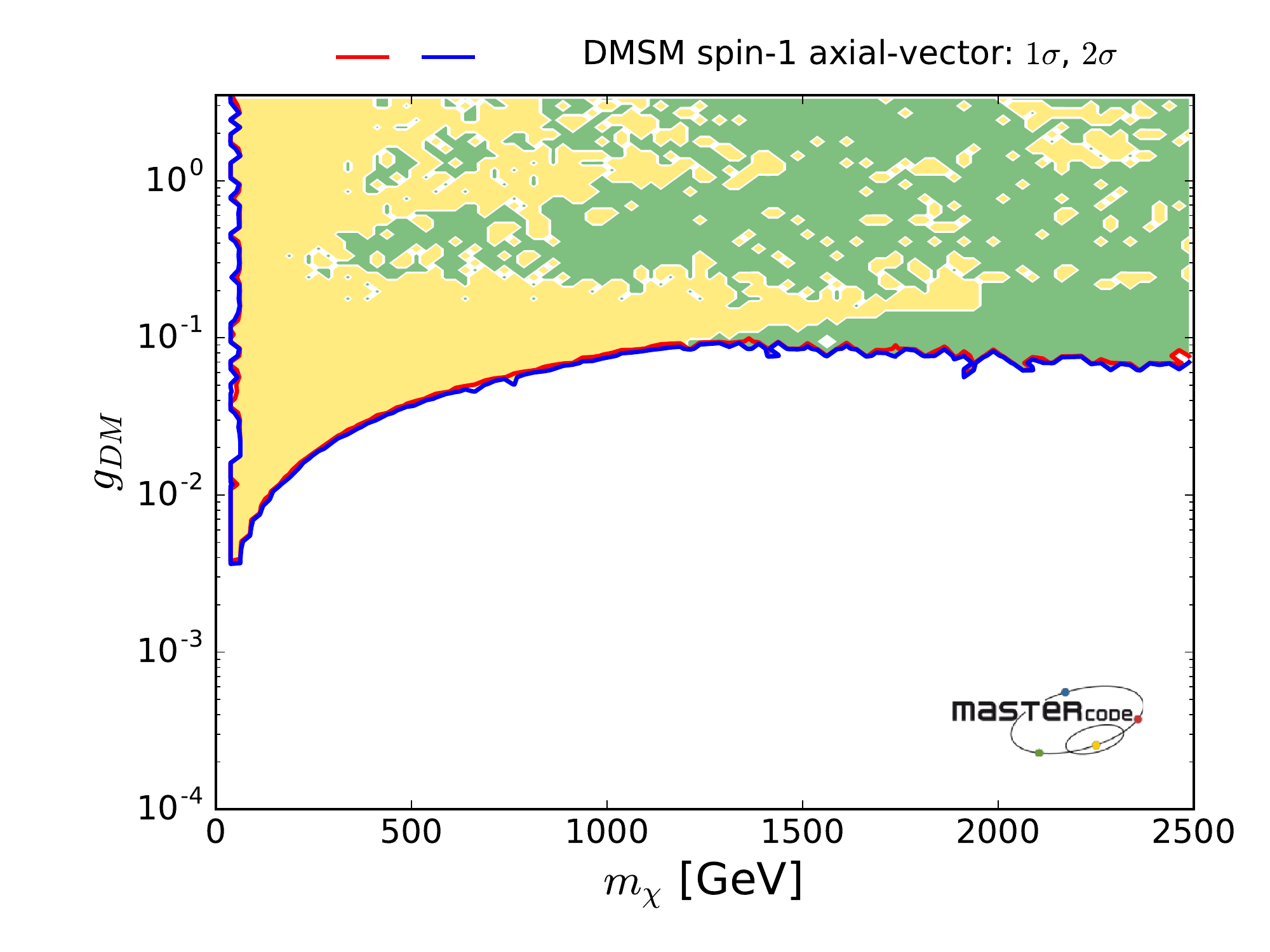}
\caption{\it The likelihood functions for $\gsm$ (left panel) and $\gdm$ (right panel)
as functions of $m_\chi$ in the model with axial-vector couplings. We again use colour coding to illustrate the
dominant mechanisms bringing the DM density into the allowed range: green for annihilation via $t$-channel
$\chi$ exchange into pairs of mediator particles $Y$ that subsequently decay into SM particles, and yellow for
rapid annihilation directly into SM particles via the $s$-channel $Y$ resonance.}
\label{fig:axialgSMandgDMvsmchi}
\end{figure*}

{The one-dimensional $\Delta \chi^2$ function for \gdm\ in the axial case
rises sharply below $\sim 4 \times 10^{-3}$, and that for \gsm\ rises above
$\sim 0.3$. The $\Delta \chi^2$ functions for $m_\chi$ and $m_Y$ are featureless above 50 and 100~GeV, respectively.}

{Finally, in Fig.~\ref{fig:ssd} we compare the experimental constraints, ranges favoured in our axial-vector DMSM
analysis and prospective experimental sensitivities to the cross section for spin-dependent scattering on a 
proton (\ssdp, inferred from the PICO-60 search with a C$_3$F$_8$ target)~\cite{PICO} (left panel) and to that
on a neutron (\ssdn, inferred from a search with the XENON1T detector)~\cite{XENONspindep} (right panel),
again accounting for the different local DM
density that we assume. {Since we consider here leptophobic mediators, the constraints
on \ssdp\ provided by Super-Kamiokande~\cite{SKnu} and IceCube~\cite{ICnu} limits on the annihilations into 
$\tau^+ \tau^-$ of DM particles trapped in the Sun are not relevant, and the limits on
hadronic annihilations are not competitive with the direct constraints on \ssdp.}
We also show in the left panel the estimated neutrino `floor' applicable to experiments using a 
C$_3$F$_8$ target {(shaded blue), 
adapted from~\cite{PICO500} using a similar factor
as in~\cite{McCabe}}.
We see that \ssdp\ approaches the PICO-60 limit most closely for
a small range $100~{\rm GeV} \lesssim m_\chi \lesssim 200$~GeV, and that \ssdp\ may be accessible to 
the LZ experiment~\cite{LZ} or the PICO-500 experiment~\cite{PICO500} for $m_\chi \lesssim 1 \tev$. However,
we see in the right panel that the most sensitive limit on spin-dependent scattering is currently set by the XENON1T
experiment, which is sensitive to \ssdn, and that the LZ experiment may be able to increase this sensitivity
significantly. {We also display the neutrino `floor' for an experiment using a Xenon target~\cite{PICO500} (shaded blue)}.
Encouragingly, we note that there are significant regions of the axial-vector
parameter space where {both \ssdp\ and \ssdn} may be detectable above the corresponding neutrino `floors',
{in both the $s$- and the $t$-channel regions}.} {However, we note that, whereas
the short-term advantage may lie with searches for \ssdn, since that currently provides
the stronger constraint, the longer-term advantage may lie with searches for \ssdp,
since the expected `floor' is lower in that case.}

.

\begin{figure*}[htb!]
\centering
\includegraphics[width=0.475\textwidth]{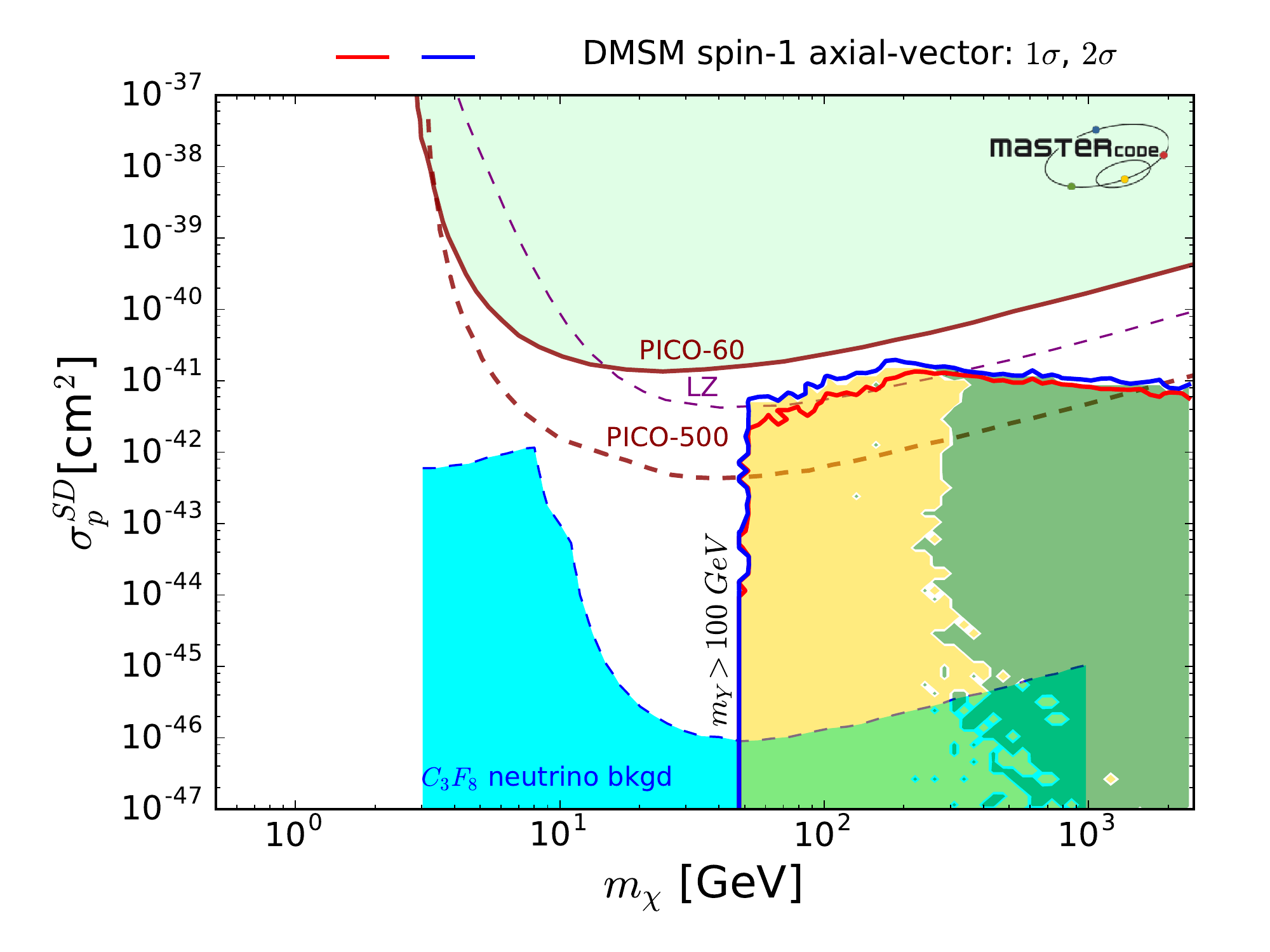}
\includegraphics[width=0.475\textwidth]{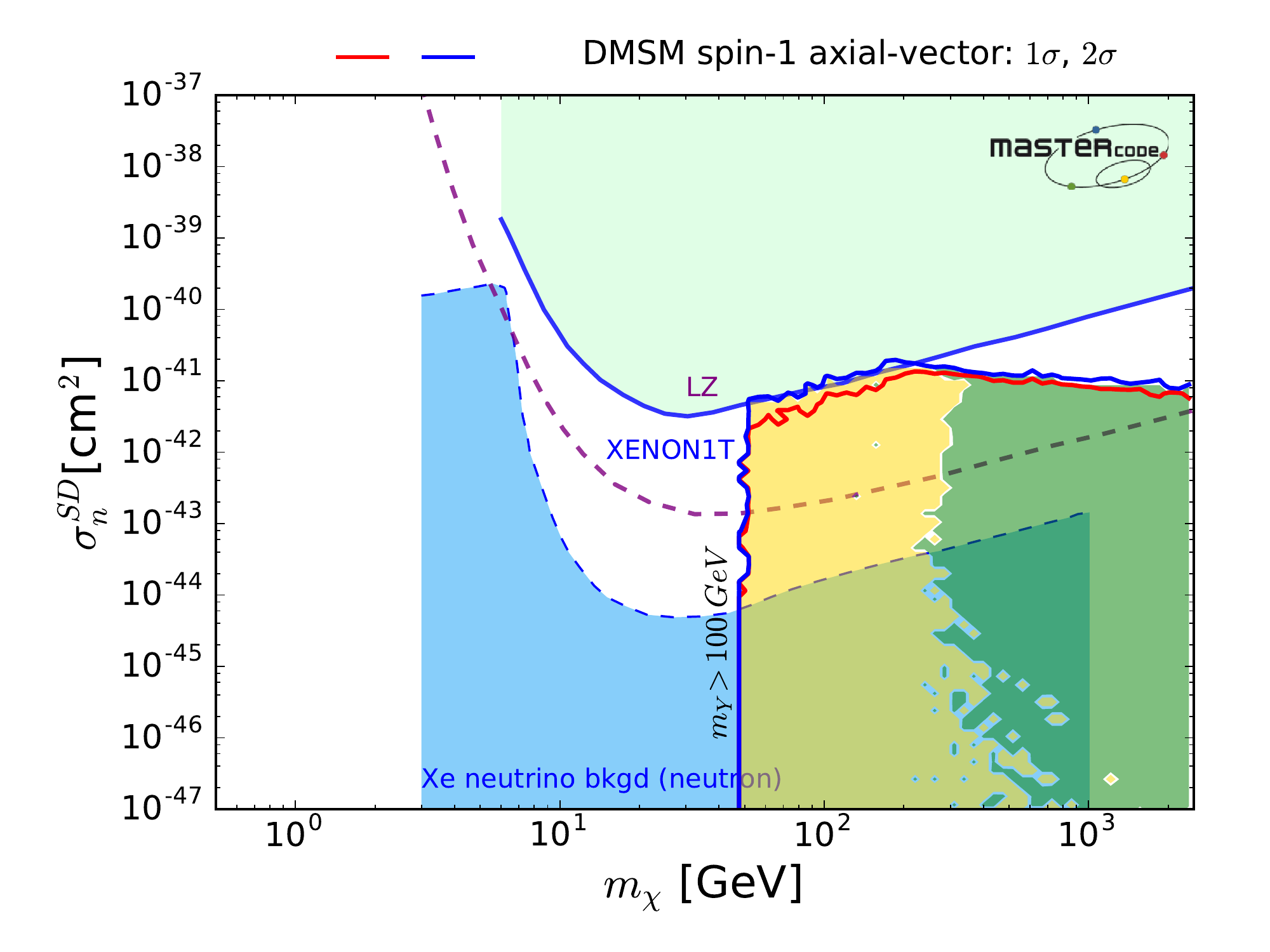}
\caption{\it {Predictions for \ssi\ and \ssdn\ in the DMSM with axial couplings.
Left panel: Contours of the likelihood function in the $(m_\chi, \ssdp)$ plane for the axial-vector model,
showing the rescaled upper limit from the PICO-60 experiment~\protect\cite{PICO} and the prospective sensitivity of the
PICO-500~\protect\cite{PICO500} and LZ~\protect\cite{LZ} experiments to \ssdp, as well as the neutrino `floor' applicable to an experiment 
such as PICO-500 that uses C$_3$F$_8$ {(shaded blue)}. Right panel: Contours of the likelihood function in the $(m_\chi, \ssdn)$ plane 
for the axial-vector model, showing the rescaled upper limit from the XENON1T experiment~\protect\cite{XENONspindep}
and the prospective sensitivity of the LZ~\protect\cite{LZ} experiment to \ssdn, {as well as the neutrino `floor' applicable to an experiment 
that uses Xenon (shaded blue)}.}}
\label{fig:ssd}
\end{figure*}




\section{Possible Ultraviolet Completions}
\label{sec:gdmeqgsm}

The vector- and axial-like leptophobic DMSMs analyzed here were chosen following the
recommendations of the LHC Dark Matter Working Group~\cite{Abdallah:2015ter,Abercrombie:2015wmb,Boveia:2016mrp,Albert:2017onk,Abe:2018bpo}, 
on the basis of their phenomenological
simplicity and without regard for their possible ultraviolet (UV) completions. In any such UV
completion, the spin-one boson could be expected to have comparable couplings to
SM and DM particles, {\it modulo} possible group-theoretical factors and mixing angles.

Looking beyond the requirements of specific grand unified or string scenarios, one should consider
the important consistency conditions on gauge couplings that are imposed by the cancellation of
anomalous triangle anomalies required for renormalizability. These entail, characteristically, that the
gauge couplings to different particle species are related by rational algebraic factors that are ${\cal O}(1)$
before mixing. Supersymmetry is an important example of a framework where such mixing factors
are important, but the couplings of supersymmetric WIMPs to the SU(2)$\times$U(1) gauge bosons
of the SM are typically not {much smaller than} those of SM particles.

The construction of DMSMs with anomaly-free U(1)$'$ gauge bosons has been studied in~\cite{Ellis:2017tkh},
where it was found that in order to be leptophobic, such DMSMs would necessarily contain non-trivial
dark sectors containing other particles besides the DM particle. Explicit examples have been given
of leptophobic DMSMs with purely vector- or axial-like couplings to quarks, and DMSMs with purely
vector- or axial-like couplings to the DM particle would be subject to further constraints. We do not
discuss the construction of such models here, but expect on the basis of the argument in the previous
paragraph that they would in general have $\gdm/\gsm = {\cal O}(1)$. Additionally,
one might expect that in a UV completion featuring unification in a non-Abelian
gauge group the spin-1 mediator couplings would be $\gtrsim {\cal O}(0.1)$, favouring parts of the
funnel regions away from the regions where $t$-channel annihilation is dominant.

We have included in Fig.~\ref{fig:vectorlogcouplingplane} and \ref{fig:axialvectorlogcouplingplane}
diagonal dotted lines where $\gdm = \gsm$ and shaded the strips where $1/3 < \gsm/\gdm < 3$, which are bounded by dashed lines.
We see that {in the vector case it} traverses the funnel region
where the DM particles annihilate via {the $s$-channel}, {and does} not approach the regions
where DM particles annihilate mainly via $t$-channel exchanges into pairs of mediator particles,
which are dominant when $\gdm \gg \gsm$. {On the other hand, in the axial-vector case both $t$- and $s$-channel annihilations are possible in the strip with $1/3 < \gsm/\gdm < 3$, as we discuss later.}
We {now discuss in more detail} the implications of the constraint $1/3 < \gsm/\gdm < 3$ for the {DMSM} parameter spaces.

We start by displaying in Fig.~\ref{fig:gVSMovergVDM} various parameter planes for the vector-coupling case
with the selection $1/3 < \gsm/\gdm < 3$.
We see that both $\gsm$ and $\gdm$ are {$> 10^{-3}$} when this
selection is imposed, and that values of $\gsm$ and {$\gdm \lesssim 0.3$} lie within the favoured range,
consistent with unification scenarios. The ranges $m_Y \gtrsim 2 \tev$
and $m_\chi \gtrsim 1 \tev$ are compatible with $\gsm, \gdm > 0.1$.}
{We note that the lower bound comes from the DM density condition, which requires $s$-channel annihilation to be fast enough. The upper bounds on $\gsm$ and $\gdm$ are, on the other hand, largely due to the spin-independent DM-nucleus scattering constraint,
with the dijet constraint also playing a role in
constraining \gsm\
for $m_Y \gtrsim 1.8$~TeV ($m_\chi \gtrsim 0.9$~TeV).}
{These features are also visible in the region of Fig.~\ref{fig:vectorlogcouplingplane} with the darker yellow band.} 

\begin{figure*}[htb!]
\centering
\includegraphics[width=0.475\textwidth]{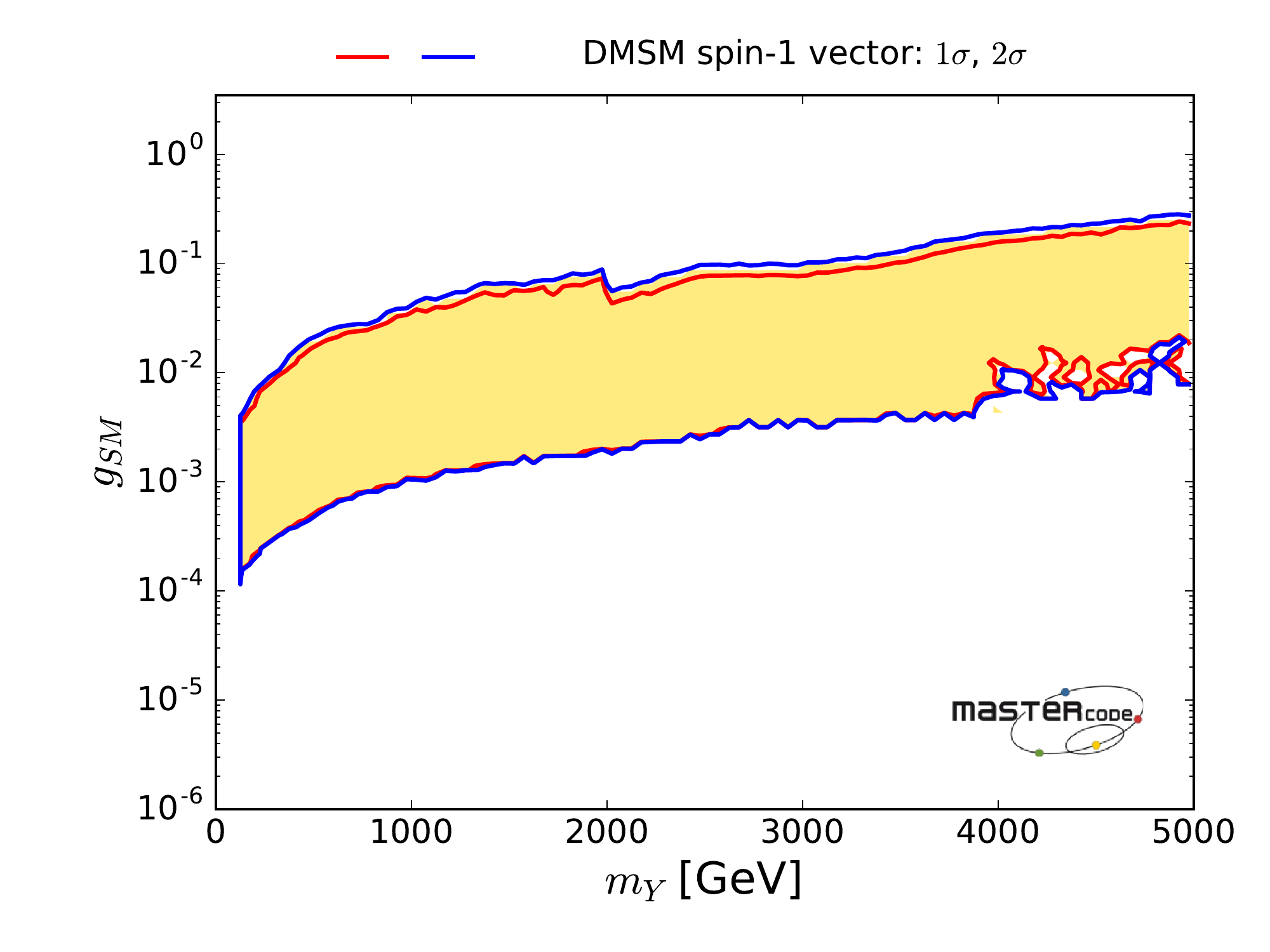}
\includegraphics[width=0.475\textwidth]{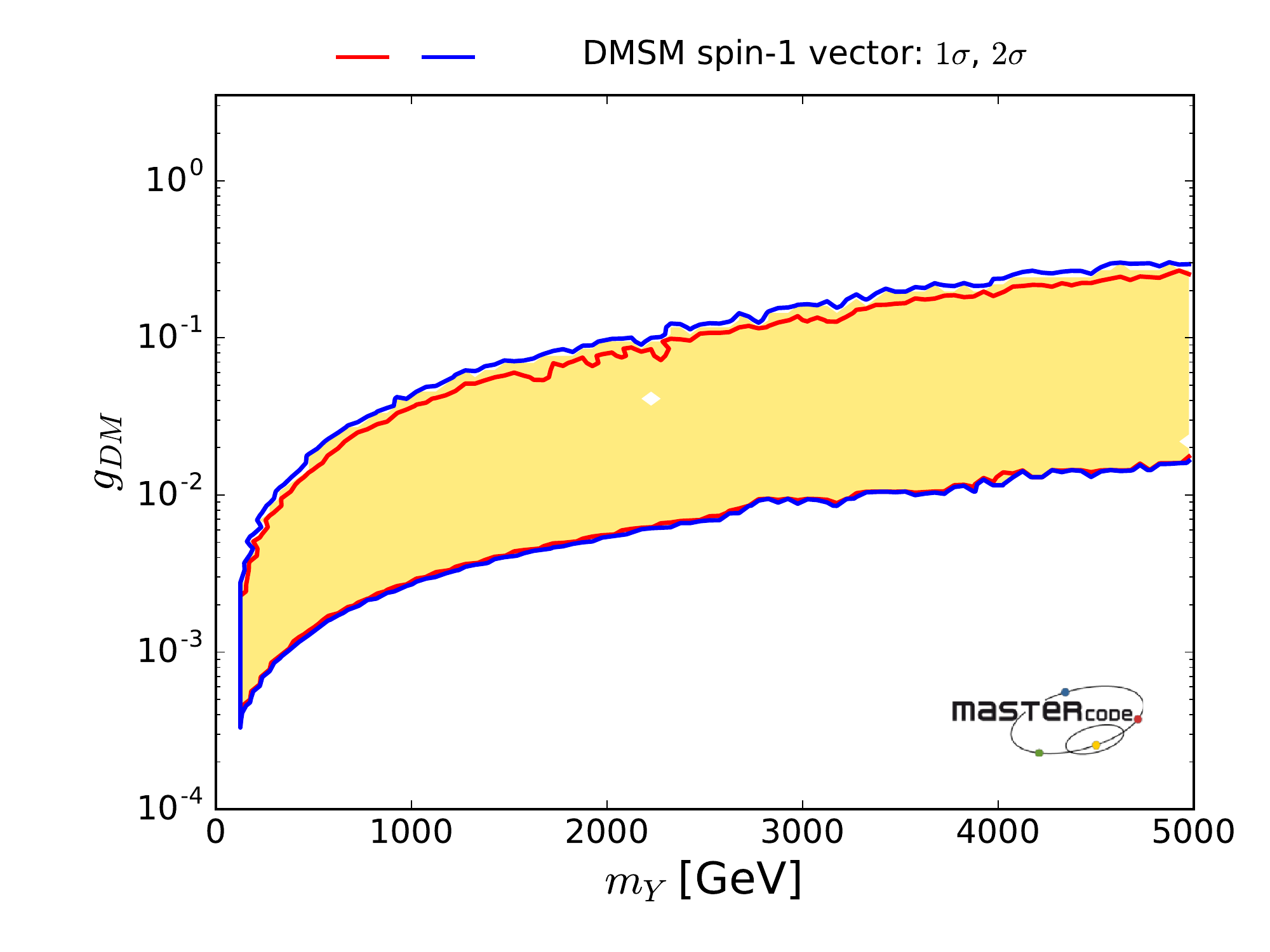} \\
\includegraphics[width=0.475\textwidth]{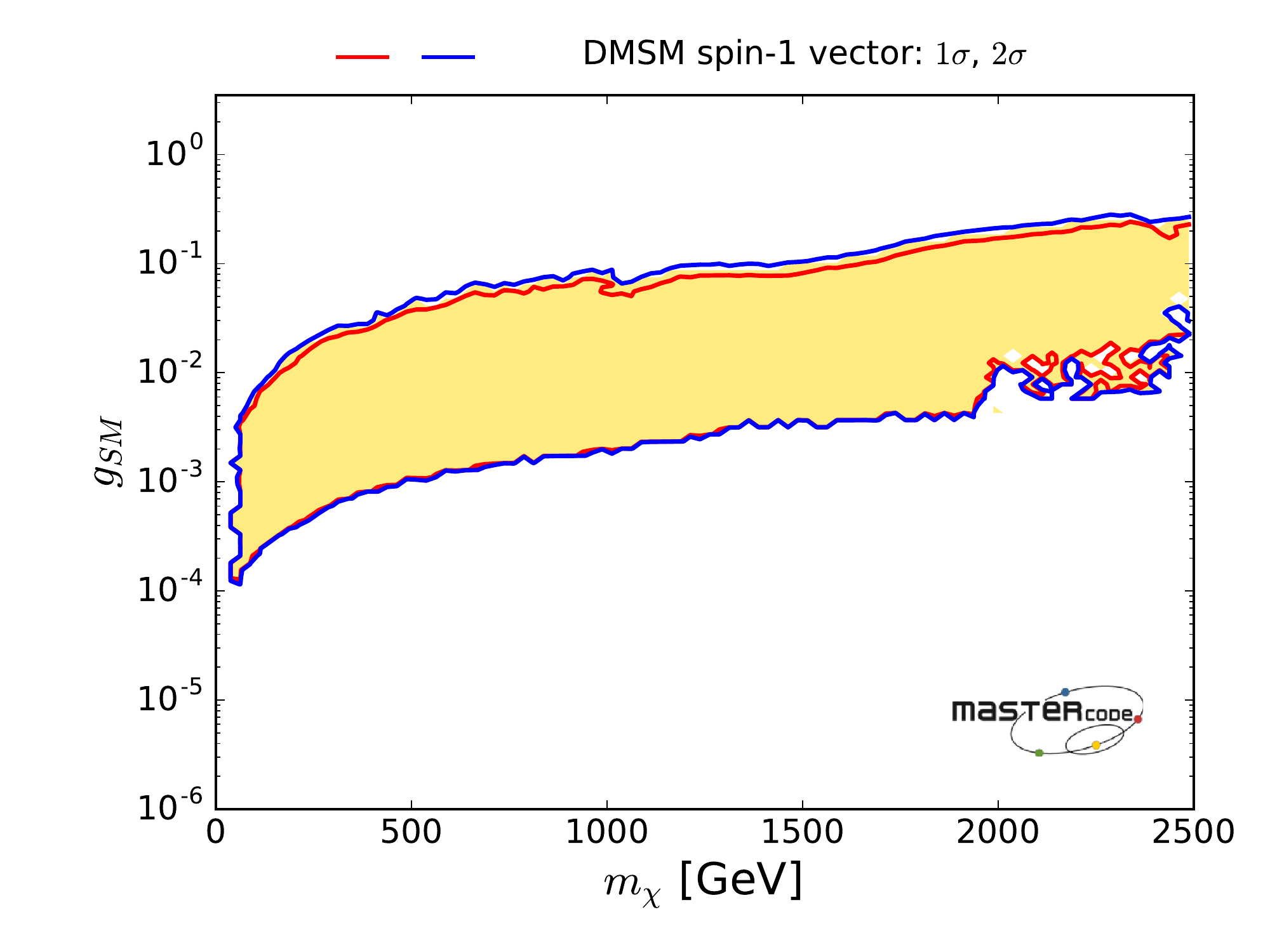}
\includegraphics[width=0.475\textwidth]{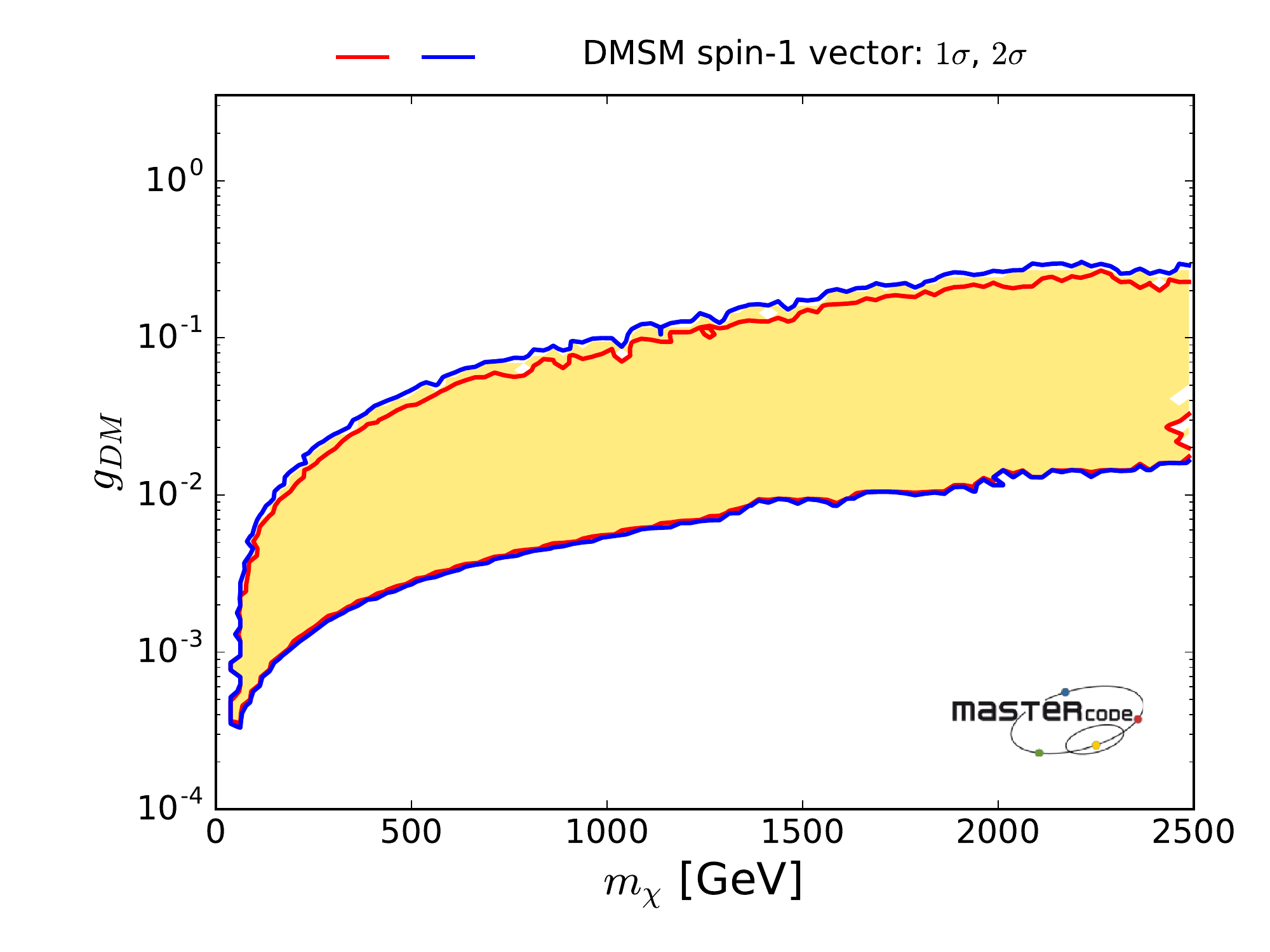}
\caption{\it The likelihood functions for $\gsm$ (left panel) and $\gdm$ (right panel)
as functions of $m_\chi$ in the model with {vector} couplings
after the selection $1/3 < \gsm/\gdm < 3$. 
We again use colour coding to illustrate the
dominant mechanisms bringing the DM density into the allowed range: green for annihilation via $t$-channel
$\chi$ exchange into pairs of mediator particles $Y$ that subsequently decay into SM particles, and yellow for
rapid annihilation directly into SM particles via the $s$-channel $Y$ resonance.}
\label{fig:gVSMovergVDM}
\end{figure*}

{Fig.~\ref{fig:gASMovergADM}} shows the corresponding parameter planes for the axial-coupling case
with the selection $1/3 < \gsm/\gdm < 3$. 
{In the upper panels, we see that the $t$-channel region is confined to a small region 
where $m_Y \lesssim 500$ GeV and {$\gsm \gtrsim 0.1$.}
{We see from the right panel of Fig.~\ref{fig:axialgSMandgDMvsmY}
that $\gdm$ must be larger than $\sim 0.3$ (0.9) at $m_Y = 500$~GeV (2~TeV) to account for the DM density.
The $1/3 < \gsm/\gdm < 3$ condition then implies that
$\gsm \gtrsim 0.1$ (0.3) for $m_Y = 500$~GeV (2~TeV).
However, such a large value of $\gsm$ is not compatible with the dijet constraint 
in this range of $m_Y$.
Therefore, no region with $m_Y \gtrsim 500$ GeV is allowed in the $t$-channel annihilation region.}
{As discussed above, $s$-channel annihilation undergoes a $p$-wave or {$m_q$} suppression in the axial-vector mediator case, 
and $\gsm \gtrsim 10^{-3}$ and $\gdm \gtrsim 3 \times 10^{-3}$ (see the darker yellow band in Fig.~\ref{fig:axialvectorlogcouplingplane}), 
are needed to compensate this suppression. However, there is a narrow band of $\gsm$ values that is also compatible with
the dijet constraint for all the sampled range of $m_Y$.
We see that the $s$-channel region is allowed for all the sampled range of $m_Y$, for the most part also if $\gsm > 0.1$.}

{In the lower panels of Fig.~\ref{fig:gASMovergADM}
we see the corresponding {$(m_\chi, \gsm)$ and
$(m_\chi, \gdm)$ planes.  
We see in the left plot that {in this projection}
the $t$-channel region appears in the coupling range $0.03 \lesssim \gsm \lesssim 0.3$,
and in the right plot its range is $0.05 \lesssim \gdm \lesssim 0.5$, 
sandwiched by the DM density condition from below and the LHC dijet constraint from above. 
{As seen in this projection},
the $t$-channel region is less restricted in the $m_\chi$ direction, and any value of $m_\chi \gtrsim 300 \gev$
is in principle {compatible with this mechanism}. 
In the $s$-channel region all values of $m_\chi$ between 50~GeV and 2.5~TeV are allowed. We note also that $\gdm, \gsm > 0.1$ are possible for
all the range of $m_\chi$.}

\begin{figure*}[]
\centering
\includegraphics[width=0.475\textwidth]{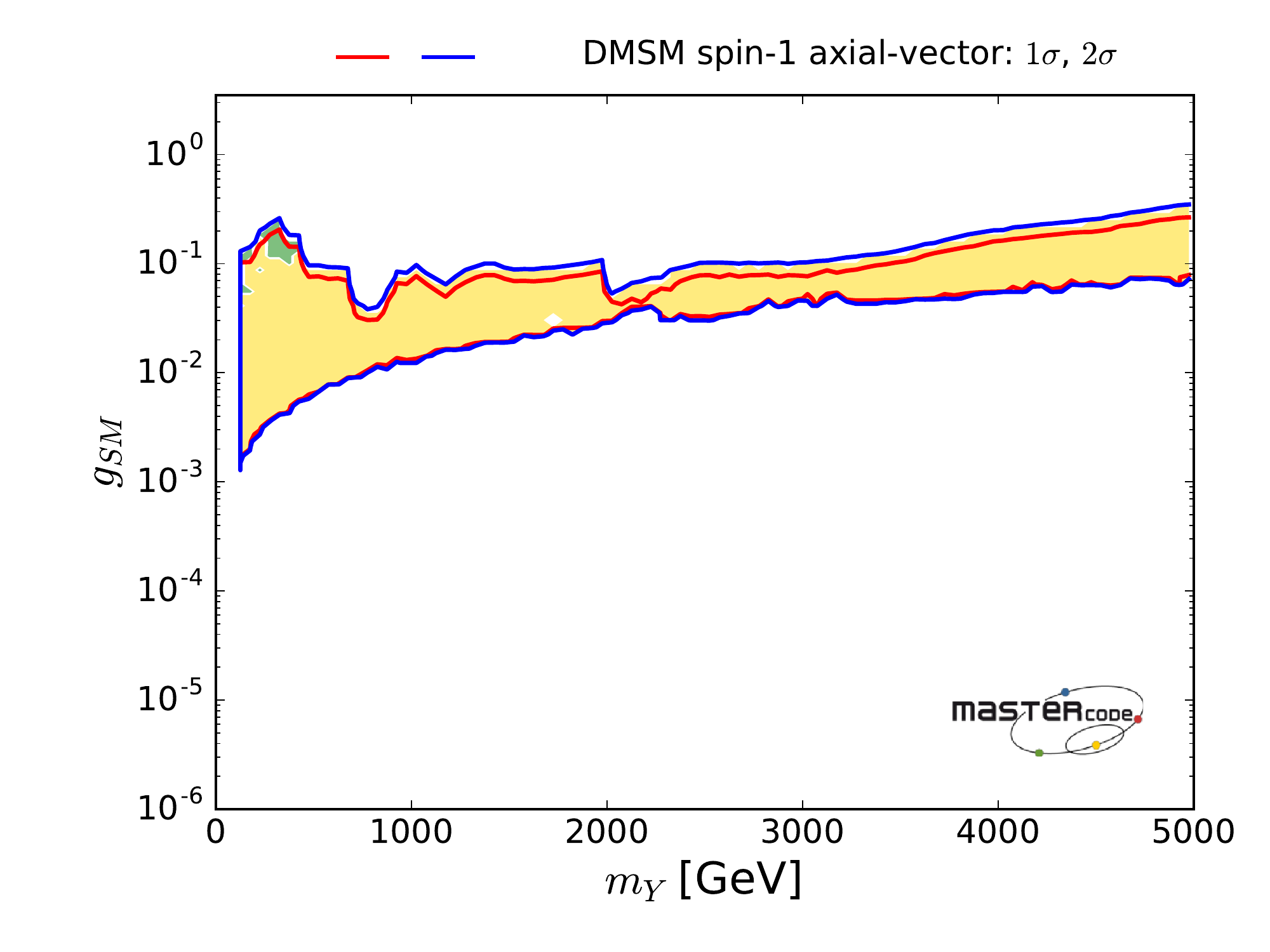}
\includegraphics[width=0.475\textwidth]{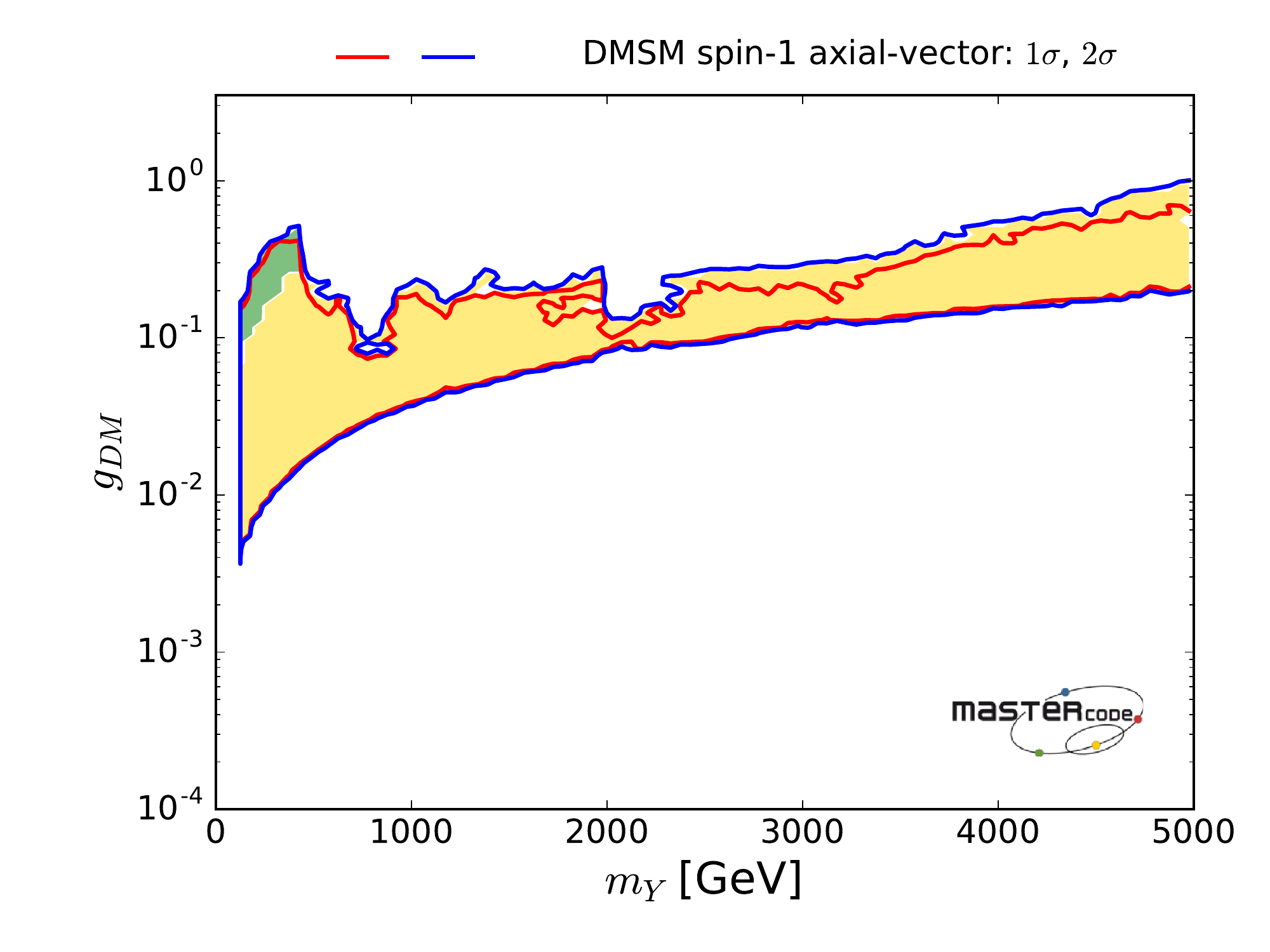} \\
\includegraphics[width=0.475\textwidth]{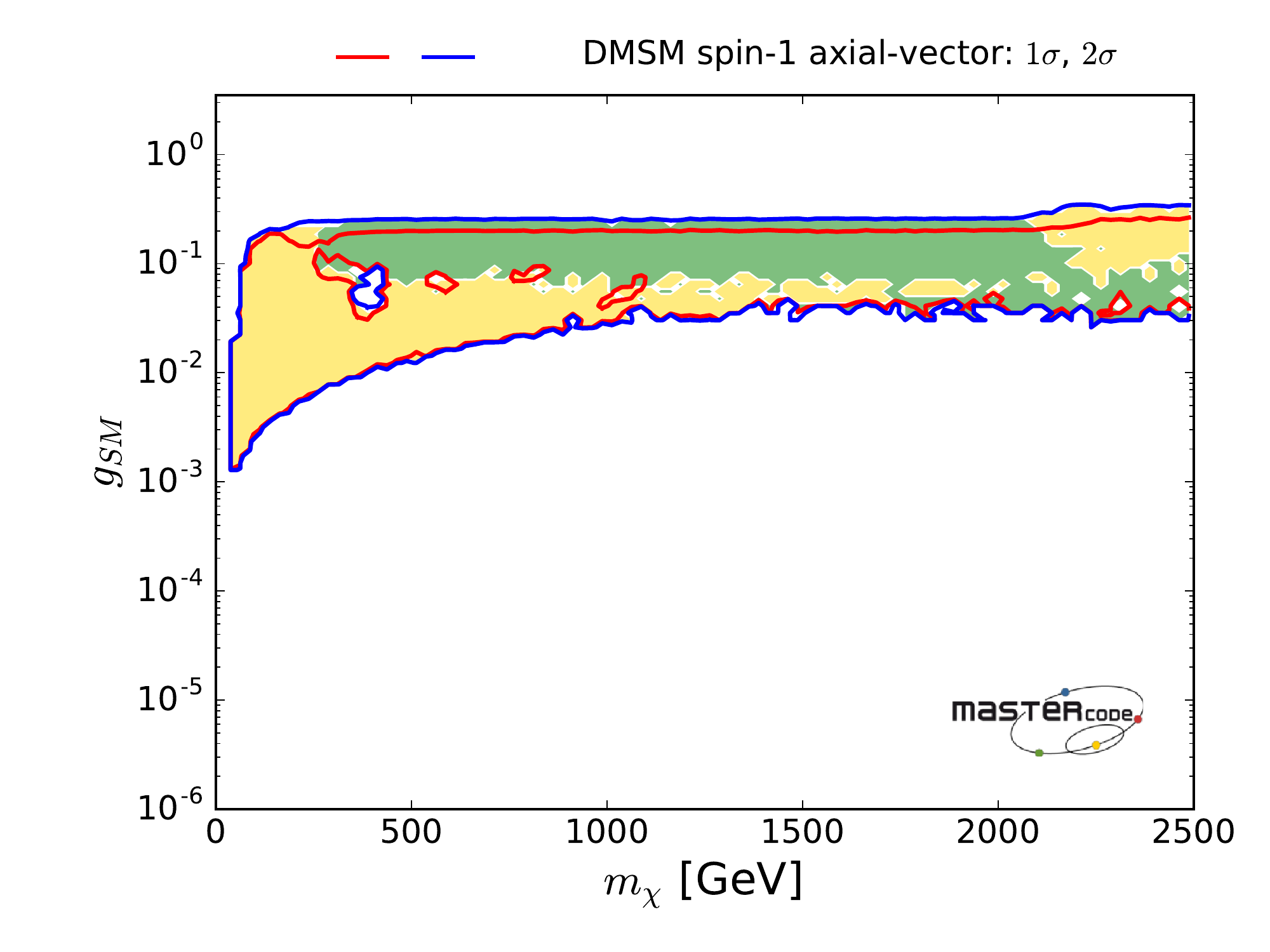}
\includegraphics[width=0.475\textwidth]{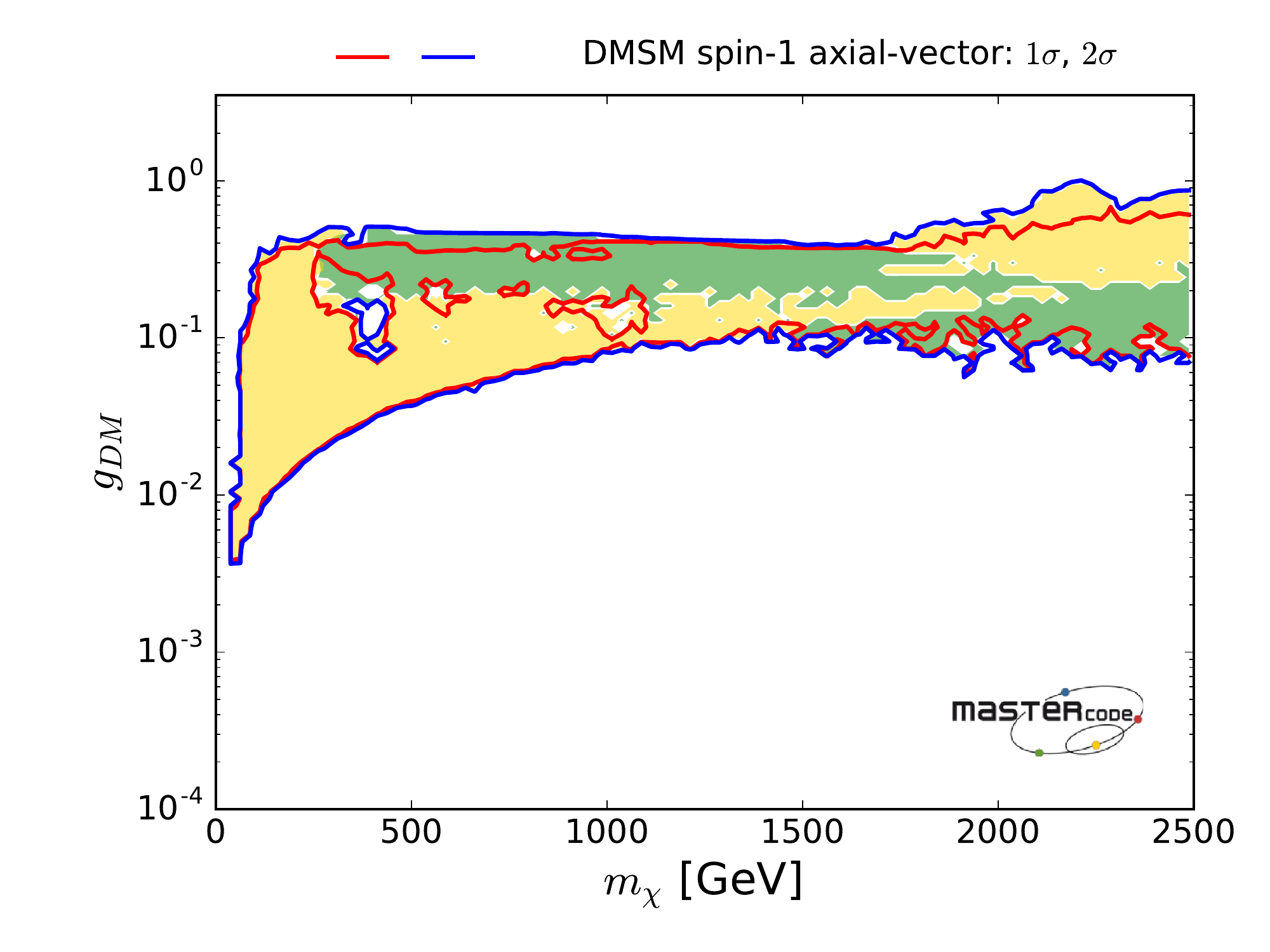}
\caption{\it The likelihood functions for $\gsm$ (left panel) and $\gdm$ (right panel)
as functions of $m_\chi$ in the model with axial-vector couplings
after the selection $1/3 < \gsm/\gdm < 3$. We again use colour coding to illustrate the
dominant mechanisms bringing the DM density into the allowed range: green for annihilation via $t$-channel
$\chi$ exchange into pairs of mediator particles $Y$ that subsequently decay into SM particles, and yellow for
rapid annihilation directly into SM particles via the $s$-channel $Y$ resonance.}
\label{fig:gASMovergADM}
\end{figure*}

{In the vector-like case, the main changes in the one-dimensional $\Delta \chi^2$ functions for \gdm\ and \gsm\ due to the selection $1/3 < \gsm/\gdm < 3$
are reductions in their upper limits to $\sim 0.3$, and there are negligible changes in the one-dimensional $\Delta \chi^2$ functions for $m_Y$ and $m_\chi$.} {In the axial case the main changes when the selection is made are that $\gdm \lesssim 0.5$ and $\gsm \gtrsim 10^{-3}$, while all values of $m_\chi \gtrsim 50 \gev$ and $m_Y \gtrsim 100 \gev$ still have very low $\Delta \chi^2$.}

{Finally, we show in Fig.~\ref{fig:gVovergDssissd} the preferred ranges of $(m_\chi, \ssi)$ in the vector-like model (left panel)
and $(m_\chi, \ssdn)$ in the axial model (right panel) after the selection $1/3 < \gsm/\gdm < 3$. We see that there are
good prospects for detecting spin-independent DM scattering in the vector-like case, since \ssi\ lies
{partially} above the neutrino `floor' within the range of $m_\chi$ sampled, though smaller values of \ssi\ could {also occur for
any value of $m_\chi > 50 \gev$.} In the case of spin-dependent scattering after the selection $1/3 < \gsm/\gdm < 3$,
we see in the right panel of Fig.~\ref{fig:gVovergDssissd}
that in the $t$-channel exchange region the predicted values of \ssdn\ are relatively close to the XENON1T limit~\protect\cite{XENONspindep},
and hence may offer prospects for detection with the
 planned LZ experiment~\cite{LZ}
within the sampled range of DM masses. However, values of \ssdn\
{may be considerably} lower in the $s$-channel exchange
region~\footnote{Note that we have again not included in the right panel 
the upper limits and prospective detection
`floors' presented by the Super-Kamiokande~\cite{SKnu} and IceCube~\cite{ICnu} Collaborations, which are based on specific assumptions
about the DM annihilation channels that are not applicable in the leptophobic DMSMs studied here.}.
This can be traced to
features visible in Fig.~\ref{fig:gASMovergADM}: we see in the upper panels that the $t$-channel region corresponds
to {small values of $m_Y$ with larger values of \gsm\ and \gdm\ than in} the $s$-channel region, and in the lower panels
{these projections show} that {the $t$-channel region may have larger values of \gsm\ and \gdm\ than in the $s$-channel region
for a large range of} values of $m_\chi$.
{The intermediate `hole' is enforced by the LHC dijet constraint, in particular.}

\begin{figure*}[htb!]
\centering
\includegraphics[width=0.475\textwidth]{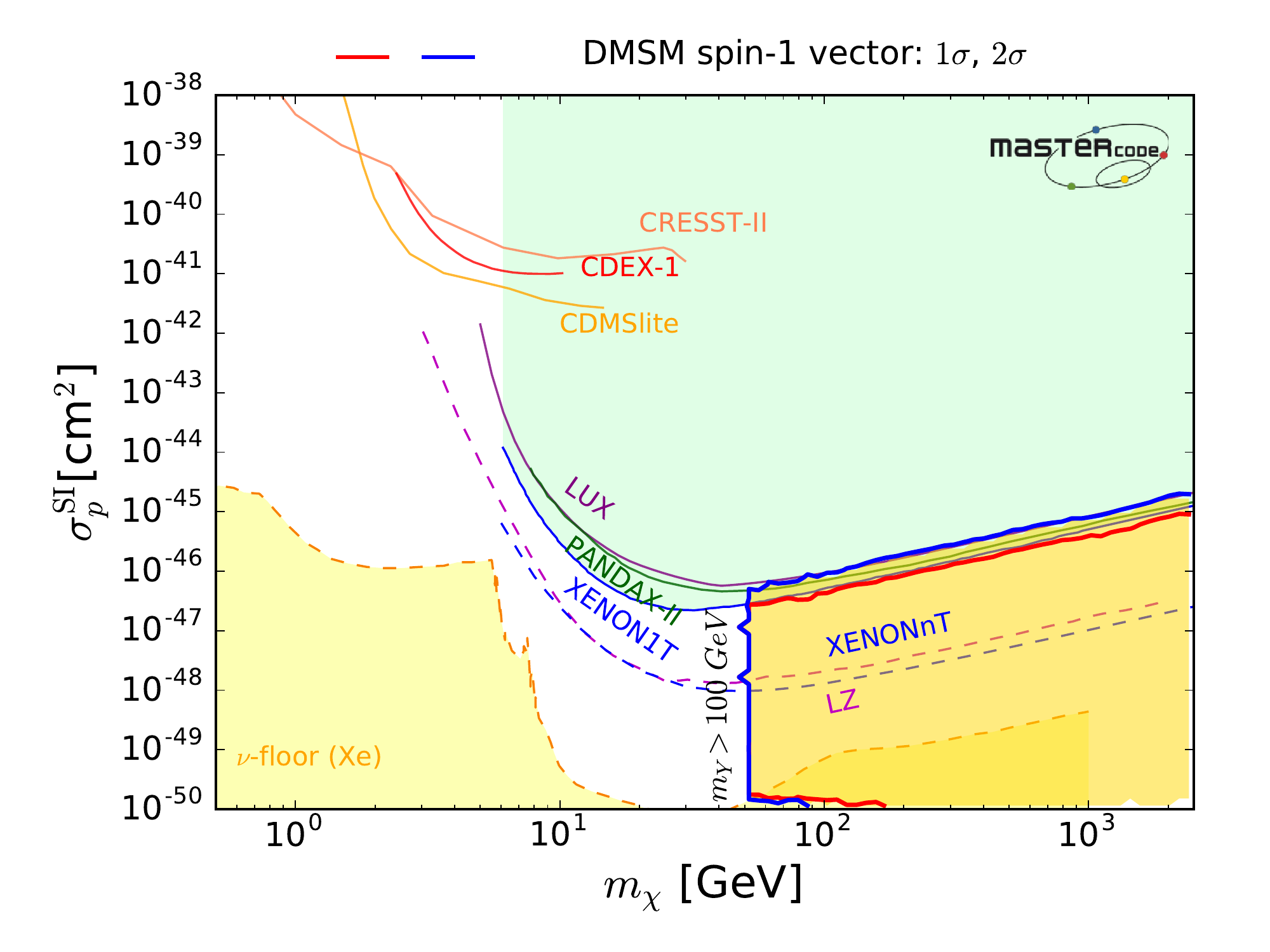}
\includegraphics[width=0.475\textwidth]{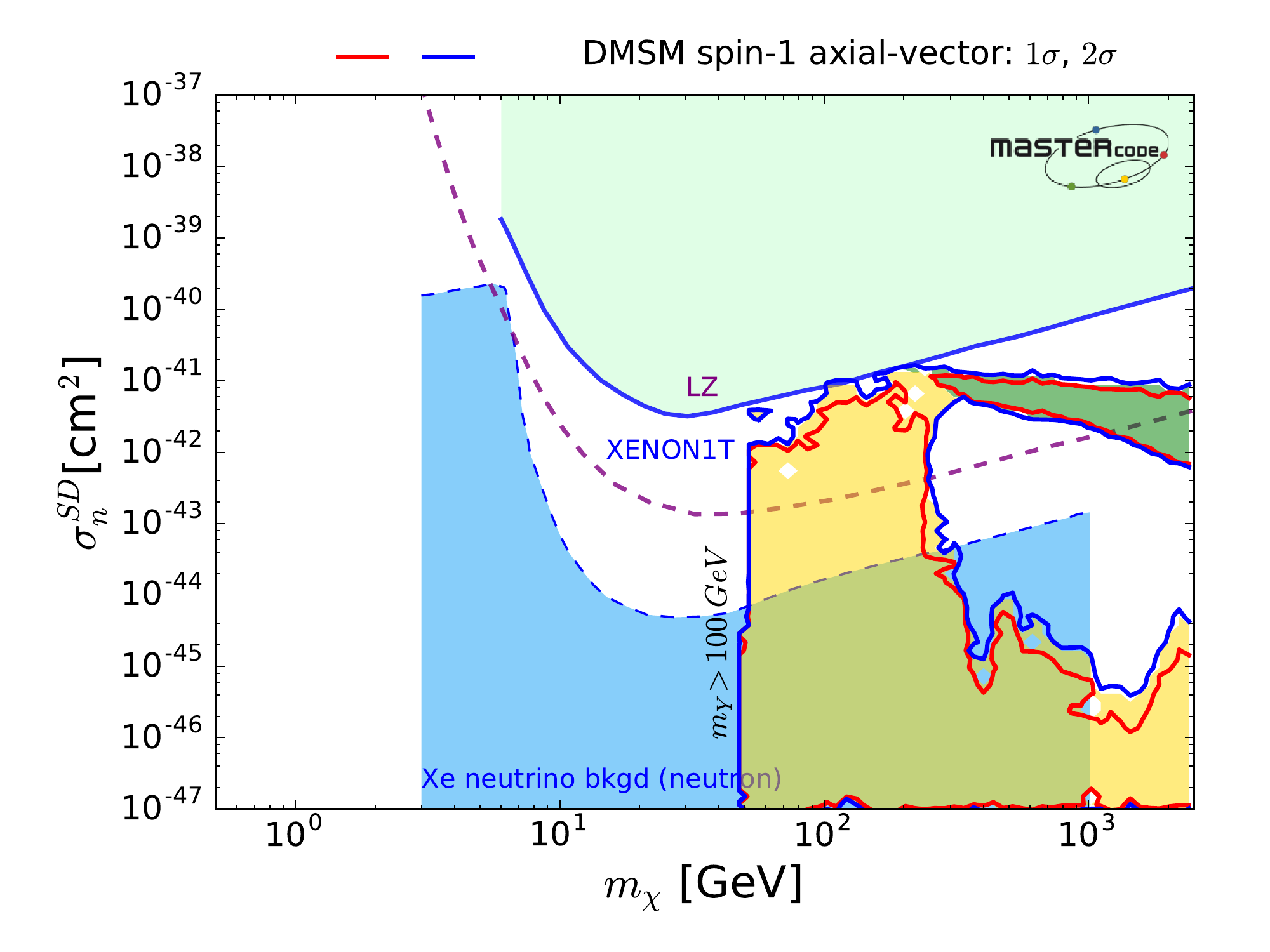}
\caption{\it {Predictions for \ssi\ and \ssdn\ after the selection $1/3 < \gsm/\gdm < 3$.
Left panel: Contours of the likelihood function in the $(m_\chi, \ssi)$ plane for the vector-like model,
showing the current upper limits from the LUX~\protect\cite{LUX}, PANDAX-II~\protect\cite{PANDAX} and
XENON1T~\protect\cite{XENON} experiments together with the neutrino `floor'~\protect\cite{floor} (shown as
the dashed orange line), and the range of \ssi\ that will be probed by the upcoming experiments
LZ~\protect\cite{LZ} and XENONnT~\protect\cite{XENONnT}. Right panel: Contours of the likelihood function in
the $(m_\chi, \ssdn)$ plane for the axial-like model,
showing the upper limit from the XENON1T experiment~\protect\cite{XENONspindep} and the prospective sensitivity of the
LZ experiment that also uses a Xenon target~\protect\cite{LZ}.}}
\label{fig:gVovergDssissd}
\end{figure*}

It is interesting that, with the selection $1/3 < \gsm/\gdm < 3$ motivated by the prospect of UV completion, 
searches for \ssi\ and \ssdn\ are {able to probe both
the $s$- and $t$-channel regions of the DMSM parameter spaces}


\section{Conclusions and Perspectives}
\label{sec:conx}

In this paper we have used {\tt MasterCode} to make a global analysis of the parameter spaces of
dark matter simplified models with leptophobic spin-one mediator particles $Y$ with either vectorial or axial couplings to
SM particles and to the dark matter particle $\chi$. Each of these models is characterized
by four free parameters: $m_Y$ and $m_\chi$, the coupling \gdm\ of the mediator to the dark matter particle,
and the coupling \gsm\ of the mediator particle to quarks, which we assume to be independent of flavour.

We have implemented constraints on the model parameter spaces due to LHC searches for
monojet events and measurements of the dijet invariant mass spectrum, as well as the cosmological
constraint on the dark matter density and  direct upper limits on spin-independent and -dependent
scattering on nuclei. We have scanned mediator masses $m_Y \le 5 \tev$ and dark matter particle
masses $m_\chi \le 2.5 \tev$, delineating the regions of the model parameters with
$\Delta \chi^2 < 2.30 (5.99)$, which are favoured at the 68\% (95\%) CL
and regarded as proxies for 1- (2-)$\sigma$ regions, respectively. Within these regions we
have identified two main mechanisms for bringing the cosmological dark matter density into the
range allowed by cosmology, namely annihilation via $t$-channel $\chi$ exchange and annihilation
via the $Y$ boson in the $s$ channel. With an eye to possible ultraviolet completions of the simplified
models studied here, we have also explored the portions of the favoured parameter space where
$1/3 < \gdm/\gsm < 3${, as discussed below}.

In the vector-like case, we find a relatively clear separation between the regions where the $t$- and $s$-channel
mechanisms dominate, with the former being more important at smaller mediator masses, small values
of \gsm\ and relatively large values of \gdm. The one-dimensional likelihood functions for both $m_\chi$
and $m_Y$ are quite {small and} flat {above} thresholds $\sim 50 \gev$ and $\sim 100 \gev$.
{Thus the LHC still has interesting prospects for discovering DM and mediator particles in these simplified models.}
Any value of \gsm\ between {$\sim 10^{-6}$ and the dijet limit of $\sim 0.3$} is possible without any $\chi^2$ penalty,
as is the case for {$\gdm \gtrsim 10^{-3}$}, where the lower limit is due to the upper limit on the sampling range
for $m_Y$. The spin-independent dark matter scattering cross section \ssi\ may be very close to the present
experimental upper limit, within the range accessible to the upcoming LZ and XENONnT experiments.
However, in both the $s$- and the $t$-channel cases \ssi\ may also be much below the neutrino `floor'.

{In the axial case the $t$- and $s$-channel regions are more} connected. The one-dimensional
likelihood functions for $m_\chi$ and $m_Y$ are again featureless above thresholds $\sim 50 \gev$ and
$100 \gev$, respectively, and that for \gsm\ is quite featureless, as is that for $\gdm \gtrsim 10^{-2}$.
We find that the spin-dependent dark matter scattering cross sections \ssdp\ and \ssdn\ may also be very close to the present
experimental upper limit, within the range accessible to the upcoming PICO-500 and LZ experiments, though
much lower values below the corresponding `floors' are also possible.

{Finally,} we have also explored the possibility that $1/3 < \gdm/\gsm < 3$, as might be suggested in some
ultraviolet completions of the dark matter simplified models considered here. In this case, values of
$\gsm, \gdm \sim 0.1$ are possible, as as might also be suggested in some scenarios with unified gauge
interactions, {and wide ranges of DM and mediator masses are again accessible to the LHC experiments}. 
In the case of vector-like couplings, this reduced range of $\gdm/\gsm$ disfavours the
$t$-channel mechanism. However, in the axial case
the $t$-channel mechanism is still possible, yields values of \ssdn\ that are relatively close to
the upper limit set by the PICO-60 experiment, and may well be within reach of the upcoming
PICO-500 and LZ experiments, whereas lower values of \ssdn\ are possible if the $t$-channel mechanism
dominates.

In this paper, we have performed a first {global study - including all the relevant {constraints from Run 2 of the LHC and direct scattering searches}
and not fixing any of the underlying parameters -}
of some DMSMs proposed by the LHC Dark Matter Working
Group~\cite{Abdallah:2015ter,Abercrombie:2015wmb,Boveia:2016mrp,Albert:2017onk,Abe:2018bpo} using the MasterCode framework. {Our analysis also provides a more detailed study of the vector and axial leptophobic models considered in~\cite{Ellis:2018xal}.} These models
are undoubtedly over-simplified, and it would be interesting and useful to extend this type of analysis to other models that
may be more realistic, which generally contain more parameters. For example, one should study models with spin-one
mediators that are not leptophobic, and also models with spin-zero mediators that may be either scalar or
pseudoscalar. One could also study models that are not flavour-universal in the quark sector, which may
be motivated by the anomalies reported in $B$ meson decays. Consistent ultraviolet completions of DMSMs
should include a mechanism for anomaly cancellation, which typically include more `dark' particles whose
possible phenomenological signatures could also be considered. The \MC\ tool is a very suitable tool for such
analyses, and we plan to use \MC\ for such analyses in the future.

\section*{Acknowledgements}
We would like to thank Y. Mambrini and M. Voloshin for 
helpful discussions.
G.W.\ acknowledges support
by the Deutsche Forschungsgemeinschaft (DFG, German Research Foundation)
under Germany's Excellence Strategy -- EXC 2121 ``Quantum Universe'' --
390833306.
 The work of K.S.\ has been partially supported by the National Science Centre, Poland,
under research grants DEC-2014/15/B/ST2/02157, DEC-2015/18/M/ST2/00054 and DEC- 2015/19/D/ST2/03136.
The work of M.B.\ and D.M.S.\ has been supported by the European Research Council via Grant BSMFLEET 639068.
The work of J.C.C.\ is supported by CNPq (Brazil). The work of M.J.D.\ is supported in part by the Australia Research Council.
The work of J.E. is supported in part by STFC (UK) via the research grant ST/L000258/1 and in part via the
Estonian Research Council via a Mobilitas Pluss grant, and the work of H.F. is also supported in part by STFC (UK) via grant ST/N000250/1.
The work of S.H.\ is supported in part by the MEINCOP Spain under contract FPA2016-78022-P, in part by the
Spanish Agencia Estatal de Investigaci\'on (AEI) and the EU Fondo Europeo de Desarrollo Regional (FEDER) through the
project FPA2016-78645-P, in part by the AEI through the grant IFT Centro de Excelencia Severo Ochoa SEV-2016-0597,
and by the "Spanish Red Consolider Multidark" FPA2017-90566-REDC. 
The work of M.L.\ is supported by XuntaGal. The work of K.A.O.\ is supported in part by
DOE grant desc0011842 at the University of Minnesota. 
During part of this work, to deploy {\tt MasterCode} on clusters we used the middle-ware suite {\tt udocker}~\cite{udocker},
which was developed by the EC H2020 project INDIGO-Datacloud (RIA 653549).
We also thank Imperial College London and the University of Bristol
for making available to us cluster computing resources that have been used intensively to carry out this work.

\end{document}